\documentclass[12pt]{article}
\usepackage{amsmath,amssymb,amsfonts,amsthm}

\newtheorem{theorem}{Theorem}
\newtheorem{lemma}{Lemma}
\newtheorem{definition}{Definition}

\newcommand{\R}{\mathbb{R}}
\renewcommand{\l}{{\cal L}}
\renewcommand{\L}{{\bf L}}

\newcommand{\loc}{\text{loc}}
\newcommand{\G}{\gamma}

\title{Initial data for fluid bodies in general relativity}
\author{Sergio Dain\thanks{Albert-Einstein-Institut, am M\"uhlenberg
    1, D-14476, Golm, Germany. E-mail: dain@aei-potsdam.mpg.de} \and
  Gabriel Nagy\thanks{D\'epartement de Math\'ematiques Facult\'e des
    Sciences, Tours, France, and the Albert-Einstein-Institut, Golm,
    Germany.  E-mail: nagy@aei-potsdam.mpg.de}}
\date{January 18, 2001}

\begin{document}
\maketitle

\begin{abstract}  

We show that there exist asymptotically flat almost-smooth initial
data for Einstein-perfect fluid's equation that represent an isolated
liquid-type body. By liquid-type body we mean that the fluid energy
density has compact support and takes a strictly positive constant
value at its boundary. By almost-smooth we mean that all initial data
fields are smooth everywhere on the initial hypersurface except at
the body boundary, where tangential derivatives of any order are
continuous at that boundary.

PACS: 04.20.Ex, 04.40.Nr, 02.30.Jr

\end{abstract}

\maketitle

\section{Introduction}

There is still missing a description by an initial value formulation
in general relativity of a self-gravitating ideal body in a general
situation. By an ideal body we mean a perfect fluid where the
thermodynamical variables and the fluid velocity have spatially
compact support. Examples are, within some approximation, a star, a
neutron star, or a fluid planet. By a general situation we mean first,
a body without symmetry, because spherically symmetric bodies are
already described, whether static \cite{aRbS91} or in radial motion
\cite{sKjE93}, and second, a situation including nearly static objects.

It is remarkable the lack of this type of description. Stars are
common objects in the universe, and a perfect fluid is the simplest
matter model for them.  General relativity is the currently accepted
theory of gravitation to describe stars, as well as planets, white
dwarfs, and neutron stars.  An initial value formulation will be a
useful tool to predict the time evolution of such objects as predicted
by Einstein's equation, without any approximation besides the choice
of the matter model.

The main difficulty is to find a solution in a neighborhood of the
time-like hypersurface corresponding to the fluid-vacuum interface,
where Einstein-Euler's equation transforms into vacuum Einstein's
equation. It is known how to describe regions not including this
interface, since an initial value formulation for vacuum Einstein's
equation was first given in \cite{yB52}, and for Einstein-Euler's
equation with everywhere non-vanishing energy density, in
\cite{yC-B58}. The problem at the interface is inherent in the fluid
equations and it is also present in a Newtonian description. A
summary of known results on free boundary problems is given in Sec.
2.5 in \cite{Rendall00}, in the context of Newton's theory as well as
of general relativity.

A first step to set up an initial value formulation for
Einstein-Euler's equation in a neighborhood of the fluid-vacuum
interface requires to find, from the complete system of equations, a
symmetric hyperbolic system that remains symmetric hyperbolic even at
that interface. A first system of this type was found in \cite{aR92}
for a certain class of fluid state functions. However, spherically
symmetric static situations can not be described by these state
functions. The reason is that every solution found in that reference
satisfies that the fluid particles at the body boundary follows a
time-like geodesic. But this is not the case for a spherically
symmetric static star, as the following argument, already given in
\cite{aR92}, shows. Consider a static spherically symmetric
stellar model. The fluid 4-velocity must be proportional to the
timelike, hypersurface orthogonal Killing vector. The proportionality
factor achieves that the 4-velocity be a unitary vector field. In the
vacuum region the space-time must coincide with Schwarzschild's. At
the star boundary the timelike Killing vector field must coincide
with the timelike Killing vector in Schwarzschild's space-time.
However, the 4-velocity obtained with this Killing vector field is
not geodesic. Therefore, spherically symmetric stellar models are not
included among the solutions given in \cite{aR92} and so, one does
not expect that nearly static stellar models can be described with
these solutions.

A second system of the type mentioned above was found, for a general
class of state functions, in \cite{hF98}. However, the initial value
formulation that describes nearly static objects is still missing.
Fix some smooth state function such that both the energy density and
the sound velocity vanish at pressure zero, and then consider the
system given in \cite{hF98} for that fluid. Assume that there exists
a smooth solution that describes such a fluid body.  Then, one can
check that the fluid particles at the boundary of the body follow a
time-like geodesics. In other words, smooth perfect fluid solutions of
system given in \cite{hF98} can not describe nearly static
situations.

Therefore, a first attempt to describe nearly static perfect fluid
bodies by an initial value formulation would be that the fluid state
function satisfies the following condition: Neither the fluid energy
density nor the sound velocity must not vanish when the pressure
vanishes. Because at the boundary of the body the pressure vanishes,
we are then requiring that the border of the body have non-zero
energy density. We call them ``liquid-type'' bodies, and
``liquid-type'' state functions. Therefore, it is natural to study
what are the appropriate initial data for a liquid-type body. It
turns out that the answer to this question was not known, it is
subtle, and is the subject of this work.

Fix once for all a simple perfect fluid, that is a perfect fluid with
1-dimensional manifold of fluid states; for example one with a state
function of the form $p(\rho)$, where $p$ is the fluid pressure and
$\rho$ the fluid co-moving energy density. Assume that this state
function is smooth and of liquid-type. By liquid-type ideal body
initial data we mean a three dimensional initial hypersurface, its
first and second fundamental forms, and the fluid initial 3-velocity
and co-moving energy density. The first two fields must be
asymptotically flat, the last two must have the compact support, the
support of the fluid velocity must be included in to the support of
the energy density; and all of them must be a solution of the
constraint equations, and satisfy some energy condition. In addition,
there exists an extra constraint on the initial fluid fields: The
fluid co-moving energy density must be {\em strictly positive and
  constant} at the border of its support. Constant because the simple
perfect fluid state function implies that there exists only one single
value of the fluid co-moving energy density such that the pressure
vanishes, and this is the value of the energy density at the border of
the body. Notice that only the energy density as measured by a fluid
co-moving observer must be constant. This extra condition only arises
for liquid-type fluids, because it is trivially satisfied in the case
where the fluid co-moving energy density vanishes at the border of the
body. [See Eqs.  (\ref{eq:fluid1})-(\ref{eq:fluid2}).]  Also notice
that this data must have a $C^1$ first fundamental form, that is, with
at least one continuous derivative everywhere; if not, Dirac's delta
appears in the fluid energy density.

Initial data of this type was not known in the literature. The only
result on solutions of the constraint equations with discontinuous
matter sources and $C^1$ first fundamental form, \cite{Choquet99},
does not guarantee that the fluid co-moving energy density be
constant at the border of the body. Here is why. The solutions with
discontinuous matter sources are found by the usual conformal
rescaling that also rescales the matter sources. Then, the initial
physical energy density [the energy density as measured by an
observer at rest with the initial surface, function $\tilde \mu$ in
Eq. (\ref{eq:fluid1})] is the product of the initial unphysical
energy density (free data) times the conformal factor at power minus
eight. We do not know any procedure to choose the free initial data
such that the solutions given in \cite{Choquet99} guarantee both,
that the fluid co-moving energy density [function $\rho$ in Eq.
(\ref{eq:fluid1})] be constant at the border of the body and the
conformal factor be $C^1$, simultaneously.  We should mention that
there are also found in \cite{Choquet99} solutions of the constraints
without rescaling the energy density, but in this case, only {\em
continuous} energy densities are considered, that is they vanish at
the border of the body.

Here, we conformally rescale all the fields except the initial
physical energy density, which is now free data, given positive and
constant at the boundary of the body. We impose that the fluid
3-velocity vanishes at the body boundary. These two conditions imply
that the fluid co-moving energy density is positive and constant at
the body boundary. [See Eq. (\ref{extra-constr}).] The subtle part
now is to solve, with the initial physical energy density as free
data, the equation for the conformal factor. This is a semi-linear
elliptic equation, with the non-linear term given by a discontinuous
function (the initial physical energy density) times the unknown (the
conformal factor) at the power plus five. In other words, a
non-decreasing function of the unknown, times a discontinuous given
function (in contrast with \cite{Choquet99} where this function is
continuous).  We introduce a compact manifold, and we prove existence
of a $C^1$ conformal factor, based on Schauder's fixed-point Theorem.
The proof follows the ideas given in an appendix of \cite{Beig96},
and in \cite{Dain99}. There is only one (technical) requirement on
the initial physical energy density: Its $L^2$-norm, computed with
the unphysical metric, must not exceed some given upper bound. We
show in the appendix that, although this condition excludes possible
initial data, it is mild enough to include interesting physical
situations, such as neutron stars.

We also give a statement on the regularity of these data. They can not
be smooth, because the liquid-type energy density is, by definition, a
discontinuous function of the space variables. How regular can it be?
The smoothest liquid-type body is almost-smooth; that is, smooth
everywhere except at the body boundary, where tangential derivatives
(appropriated defined) of any order are continuous. In other words, we
prove the following: If the unphysical metric is a smooth field on the
initial hypersurface, and the fluid free initial data are smooth up to
the body boundary with every derivative tangential to the boundary
continuous through that boundary, then the same holds for all the
initial data fields. A crucial requirement to prove this statement is
that the conformal factor be $C^1$ on a neighborhood of the body
boundary. (See Sec. \ref{s-reg}.)

Summarizing, we prove that {\em almost-smooth initial data
representing liquid-type simple perfect fluid bodies can be obtained
as solutions of the constraint equations, in the case where the
fluid 3-velocity vanishes at the body-boundary, through a suitable
modification of the usual conformal rescaling techniques.}

Finally, some technical remarks: (i) We rescale the initial fluid
momentum density but not the initial fluid energy density, in order to
solve the constraint equations. Therefore, we have to choose the
rescaled momentum density small enough, in the sense given in Sec.
\ref{EC}, in order to have physical data satisfying the dominant
energy condition.  (ii) We impose that the fluid 3-velocity vanish at
the body boundary.  There is no physical justification for this
assumption, it is made because is the only way we know, with the
rescaling of the initial data field that we have chosen, to guarantee
that the fluid co-moving physical energy density be constant at the
body boundary. [See Eq. (\ref{extra-constr}).]  We give an
interpretation of this condition in Sec. \ref{S:MR}.  (iii) Besides
the conformal rescaling to solve the constraint equations we perform a
conformal compactification in order to solve elliptic equations on
some unphysical compact manifold. Asymptotic decay properties of
fields in the physical initial hypersurface are translated into
differentiability properties of these fields at a particular point in
the unphysical compact manifold. These differentiability properties at
the point at infinity are completely independent of the
differentiability of the fields near the body boundary.

In Sec.  \ref{s-dmr} we introduce the main definitions we need in
order to present the principal result, Theorems \ref{T:main1} and
\ref{T:main2}. We also give a proof based on results obtained in
Secs. \ref{s-er} and \ref{S:FR}. In Secs. \ref{s-exist} and
\ref{SS:MC} we give the main existence proofs for the semi-linear and
linear elliptic equation associated with the Hamiltonian and momentum
constraint, respectively. In Sec. \ref{s-reg} we prove the regularity
statements, Theorems \ref{T:p-smooth} and \ref{T:t-smooth}. In Sect
\ref{EC} we explain why this discussion on an energy condition
appears, and we give a simple condition on the free data such that
the physical initial data satisfies the dominant energy condition.
The constraint equations are naturally written in terms of the
initial fluid energy density and the initial fluid momentum density.
Eqs. (\ref{eq:fluid1})-(\ref{eq:fluid2}) relate them to the fluid
pressure, co-moving energy density, and 3-velocity.  In Sec. 
\ref{s-Bw}, we prove that these equations are invertible.  In Sec. 
\ref{s-d}, we comment on the initial value formulation for
liquid-type ideal bodies. In Theorem \ref{T:Hc-exist1} we need to
assume that the fluid initial energy density satisfies an inequality
[Eq. (\ref{eq:b-vtheta})] involving both the unphysical manifold and
the unphysical rescaled metric. In Appendix \ref{s-a}, we study this
inequality. In Sec. \ref{s-tc}, we show (Lemma \ref{L:gen-bound})
that a similar (but weaker) inequality holds for every initial data.
By an explicit example, in Sect \ref{s-sp}, we prove that the
inequality required for the existence theorems is in fact a
restriction on the allowed initial data. This example also suggests
that this restriction is mild, in the sense that interesting physical
systems, like neutron stars, satisfy it.

\section{Definitions and main result}
\label{s-dmr}

\subsection{Liquid-type ideal body data}
\label{liquidtype}

We first introduce what we mean by initial data for a liquid-type
ideal body. Afterwards, we split the concept of almost-smooth into
two pieces. Given a field and an open bounded set $\Omega$ on some
manifold, we introduce the concept of an $\Omega$-piecewise smooth and
$\Omega$-tangentially smooth field. Finally, in the next subsection,
we present our main result.

Consider an initial data set for Einstein's equation with matter. That
is, consider a 3-dimensional, smooth, connected manifold $\tilde{M}$,
a positive definite metric, $\tilde{q}_{ab}$, and a symmetric tensor
field, $\tilde{p}^{ab}$, on $\tilde{M}$, together with a vector field,
$\tilde{j}^a$, and a positive scalar function, $\tilde{\mu}$, subject
to the condition $\tilde{j}_a\tilde{j}^a \leq \tilde{\mu}^2$, and
solution on $\tilde{M}$ of
\begin{align}
\label{eq:constr1}
\tilde D_a \tilde p^{ab} - \tilde D^b \tilde p_a{}^a 
&= - \kappa \tilde j^b, \\
\label{eq:constr2}
\tilde R + (\tilde p_a{}^a)^2 - \tilde p_{ab}\tilde p^{ab} 
&=  2\kappa \tilde \mu,
\end{align}
where $\tilde{D}_a$ and $\tilde R$ are the Levi-Civita connection and
the Ricci scalar associated with $\tilde{q}_{ab}$, and $\kappa = 8
\pi$. Indices on tensors with ``tilde'' are raised and lowered with
$\tilde q^{ab}$ and $\tilde q_{ab}$, respectively, where $\tilde
q_{ac} \tilde q^{cb} = \delta_a{}^b$.  Latin letters $a$, $b$, $c$,
represent abstract indices.  The fields solving
(\ref{eq:constr1})-(\ref{eq:constr2}) have a meaning as part of a
4-dimensional space-time solution of Einstein's equation with matter
sources. The manifold $\tilde{M}$ represents a three dimensional
space-like hypersurface such that $\tilde{q}_{ab}$ and
$\tilde{p}^{ab}$ are their first and second fundamental forms. This
hypersurface will be a maximal slice if and only if $\tilde p_a{}^a
=0$. The fields $\tilde \mu$ and $\tilde{j}^{a}$ represent the
normal-normal and the (negative) normal-parallel components to $\tilde
M$ of the stress-energy tensor. The dominant energy condition on the
stress-energy tensor in the space-time implies $\tilde{j}_a
\tilde{j}^a \leq \tilde{\mu}^2$ on $\tilde M$. This condition is the
reason why one does not, in general, pick {\em any} $\tilde q_{ab}$
and $\tilde p^{ab}$ and then {\em define} $\tilde j^a$ and $\tilde
\mu$ by (\ref{eq:constr1})-(\ref{eq:constr2}). Because if one does
that, the resulting fields form an initial data set iff the energy
condition is satisfied by these $\tilde j^a$ and $\tilde \mu$. It is
an open question whether there exists a procedure to find appropriate
$\tilde q_{ab}$ and $\tilde p^{ab}$ besides the conformal rescaling
one.

The initial data set is asymptotically flat if the complement of a
compact set in $\tilde M$ can be mapped by a coordinate system
$\tilde x^j$ diffeomorphically onto the complement of a closed ball
in $\R^3$ such that we have in these coordinates
\begin{equation} 
\label{eq:pf1}
\tilde q_{ij}=\left(1+2m/\tilde r\right)\delta_{ij}+O(\tilde r^{-2}),
\end{equation}
\begin{equation} 
\label{eq:pf2}
\tilde p^{ij}=O(\tilde r^{-2}),
\end{equation}
as $\tilde r:= \sqrt{\delta_{ij}\tilde x^i\tilde x^j} \to \infty$,
where $m$ is a constant that represents the ADM mass of the
data. Latin letters $i$, $j$, $k$, denote coordinates indices and take
values 1, 2, 3, while $\delta_{ij}= \mbox{diag}(1,1,1)$.

Fix as matter source a simple perfect fluid. That is, first,
introduce on $\tilde M$ a non-negative scalar field $\rho$,
interpreted as the fluid co-moving energy density, a vector field
$\tilde v^a$, interpreted as the fluid initial 3-velocity, and fix a
function $p(\rho)$, the state function, interpreted as the fluid
pressure as function of the co-moving energy density. Second,
introduce on $\tilde M$ the equations
\begin{align}
\label{eq:fluid1}
\tilde \mu &= \frac{\rho+p \tilde v^2}{1-\tilde v^2},  \\
\label{eq:fluid2}
\tilde j^b &= \frac{(\rho+p)\tilde v^a}{1-\tilde v^2},
\end{align}
with $\tilde v^2 = \tilde v_a \tilde v^a < 1$. In the space-time
solution of Einstein's equation with matter sources, the normal-normal
and the (negative) normal-parallel components to $\tilde M$ of the
usual perfect fluid stress-energy tensor are precisely the left hand
side of Eqs.  (\ref{eq:fluid1})-(\ref{eq:fluid2}), respectively.  If
$u^a$ denotes the unit fluid 4-velocity, and $n^a$ the unit normal to
the initial hypersurface, then $u^a=(n^a+\tilde v^a)/\sqrt{1-\tilde
  v^2}$. For this matter model the dominant energy condition is
equivalent to $\rho \geq p$. In addition one can prove that $\rho \geq
p$ implies $\tilde{j}_a\tilde{j}^a < \tilde{\mu}^2$.  The sketch of
the proof is the following: from Eqs.
(\ref{eq:fluid1})-(\ref{eq:fluid2}) define $f(\rho,\tilde v) :=
\sqrt{\tilde j_a\tilde j^a}/ \tilde \mu$. Notice that $f(\rho, 0) =0$,
and $f(\rho,1) =1$. One can prove that $\rho \geq p$ implies  
$\partial f/\partial \tilde v > 0$, for $0 \leq \tilde v < 1$; then,
follows that $f<1$ for  $0 \leq \tilde v < 1$. 

Consider a liquid-type ideal body data set. That is, an
asymptotically flat initial data with a simple perfect fluid whose
state function is of liquid-type, and both the fluid 3-velocity and
the co-moving energy density have the same compact support,
$\overline{\Omega} \subset \tilde M$. By a liquid-type state function
we mean a non-negative, non-decreasing, smooth function $p(\rho)$
that vanishes at $\rho_0 >0$. As an example consider a big water drop
(in order to neglect surface tension effects), or a fluid planet, or a
neutron star.  The support of the energy density represents the place
occupied by the body. The value of the co-moving energy density at
the border is determined by the function of state as the value where
the pressure vanishes. (Otherwise, the acceleration of fluid
particles lying on this border becomes infinite.)  Therefore, a
liquid-type ideal body satisfies $\rho|_{\partial \Omega} = \rho_0$.
Eqs. (\ref{eq:fluid1})-(\ref{eq:fluid2}) translate this condition for
the co-moving energy density into a constraint on the fields $\tilde
\mu$ and $\tilde j^a$ at $\partial \Omega$, where they are not longer
free but they must satisfy
\begin{equation}
\label{extra-constr}
\left. \left[\tilde \mu
\left(1 - \tilde j_a\tilde j^a/\tilde \mu^2\right)
\right]\right|_{\partial \Omega} = \rho_0.
\end{equation}

As a summary, we state the following:
\begin{definition}
\label{d-lt-fluid}
A liquid-type ideal body initial data set consists of fields $\tilde
q_{ab}$, $\tilde p^{ab}$, $\tilde v^a$, and $\rho$ on $\tilde M$, and
a state function $p(\rho)$, such that: (i) $\tilde q_{ab}$ is a
Riemannian metric, $\tilde p^{ab}$ is a symmetric tensor, and both
are asymptotically flat; (ii) $p(\rho)$ is liquid-type, and vanishes
at $\rho_0>0$; (iii) $\mbox{supp}(\tilde v^a)\subset \mbox{supp}(\rho) =
\overline\Omega$ compact; (iv)  $\rho|_{\partial \Omega} = \rho_0$;
(v) These fields are solutions of Eqs. 
(\ref{eq:constr1})-(\ref{eq:constr2}),
(\ref{eq:fluid1})-(\ref{eq:fluid2}) on $\tilde M$.
\end{definition}

Given an open set $\Omega' \subset \R^3$, we denote by
$C^{s}(\Omega')$ and $C^{s,\alpha}(\Omega')$ the spaces of $s$-times
continuously and H\"older continuously differentiable functions,
respectively, with $s\geq 0$ integer, and $0 < \alpha < 1$. We use
the notation $C^{\alpha}(\Omega') = C^{0,\alpha}(\Omega')$. We also
denote by $L^p(\Omega')$, $W^{s,p}(\Omega')$, and by
$L^p_{\loc}(\Omega')$, $W^{s,p}_{\loc}(\Omega')$ the Lebesgue and
Sobolev spaces, and the local Lebesgue and local Sobolev spaces,
respectively, where $1 < p < \infty$. We follow the definitions given
in \cite{Adams,Gilbarg}, and the generalizations for smooth 
manifolds, $M'$, given in \cite{Aubin82}. Finally, we say that a
tensor field on $\Omega'\subset M'$ belongs to one of the functional
spaces mentioned above, if all its components, in some smooth atlas of
$M'$, belong to such a space.

It is convenient to split the concept of an almost-smooth field,
presented in the introduction, into the following two definitions.
The first one is an $\Omega$-piecewise smooth field. Consider a
smooth manifold $M'$, a tensor field $u$ on that manifold, and a
 open set $\Omega \subset M'$, with compact closure. We say
that $u$ is $\Omega$-piecewise smooth, if $u \in C^{\infty}(\overline
\Omega) \cap C^{\infty}(M'\setminus \Omega)$. Note that this
definition involve conditions on the field both in $\Omega$ and its
complement.  An example of an $\Omega$-piecewise smooth but not
smooth function is any $f$ such that, $0 < f\in C^{\infty}(\overline
\Omega)$ and $f =0$ on $M' \setminus \overline \Omega$. The fluid
energy density of a liquid-type body is such a function.

The second concept is an $\Omega$-tangentially smooth field.  Let
$q_{ab}$ a smooth, positive definite, metric on $M'$. Assume that
$\partial \Omega$ is a smooth submanifold of codimension one. Let
$\hat n^a$ a normal vector to $\partial \Omega$ with respect to
$q_{ab}$. Consider a Gaussian normal foliation in a neighborhood of
$\partial \Omega$, that is, a foliation orthogonal to the geodesics
tangent to $\hat n^a$ at every point of $\partial \Omega$. Define
$\hat n^a$ outside $\partial \Omega$ to be tangent to these geodesics.
Let $V_{\partial \Omega}^a$ any smooth tangent vector field to this
foliation, i.e.; any smooth vector field such that $V_{\partial
  \Omega}^a\hat n_a=0$.  We say that an $\Omega$-piecewise smooth
field $u$ is $\Omega$-tangentially smooth if for all $k\geq 1$ the
tangential derivatives $V^{(k)}_{\partial \Omega}(u)$ are continuous;
where $V^{(1)}_{\partial \Omega}(u) := V^a_{\partial \Omega}D_au$, and
$V^{(k)}_{\partial \Omega}(u) := V^a_{\partial
  \Omega}D_a[V^{(k-1)}_{\partial \Omega}(u)]$, for $k\geq 1$.  For
example, choose the field to be the energy density of a liquid-type
ideal body, and $\Omega$ the interior of its support. A necessary
condition for  this field to be $\Omega$-tangentially smooth is to
be constant at $\partial \Omega$.

\subsection{Main result}
\label{S:MR}

The strategy is, first, to find fields $\tilde q_{ab}$, $\tilde
p^{ab}$, $\tilde j^a$, and $\tilde \mu$, solution of
(\ref{eq:constr1})-(\ref{eq:constr2}) with the desired properties.
Conformal rescaling techniques are used in this part. We also
introduce a compact manifold where equations associated with
(\ref{eq:constr1})-(\ref{eq:constr2}) are solved with boundary
conditions chosen in such a way that, the de-compactification of these
solutions gives asymptotically flat initial data. Then we prove that
under specific assumptions on the state function, Eqs.
(\ref{eq:fluid1})-(\ref{eq:fluid2}) can be inverted for all $\rho \geq
\rho_0$ and for $\tilde v^a$ with $0 \leq \tilde v < 1$.

Fix a 3-dimensional, orientable, connected, compact, smooth manifold.
Fix $i \in M$, and $\tilde M := M\setminus \{i\}$. The choice $M=S^3$,
and so $\tilde M= \mathbb{R}^3$,  describes, for example, ordinary
stars.  A restriction of this type in the topology, however, plays no
role in what follows.  Let $h_{ab} \in C^{\infty}(M)$ be a Riemannian
metric on $M$. Let $x^i$ be its associated Riemann normal coordinate
system at $i$, and $r$ the geodesic distance. Let $\hbar_{ab} \in
C^{\infty}(M)$ a symmetric tensor such that
\begin{equation}
  \label{eq:15}
  x^i\hbar_{ij}=0.
\end{equation}
Latin indices $i,j,k$ denote tensor components on coordinates $x^i$.
Let $q_{ab} \in C^{\infty}(\tilde M)$, be a Riemannian metric on $M$
with scalar curvature $R$.  Let $B_{\epsilon}$ be an  open ball of
geodesic radius $\epsilon$ centered at $i$.  Assume that there exists
$\epsilon >0$, such that the metric $q_{ab}$  on $B_{\epsilon}$ has
the form
\begin{equation}\label{A:metric}
q_{ij} =h_{ij} + r^3 \hbar_{ij},
\end{equation}
in the coordinates $x^i$. Since we have assumed (\ref{eq:15}), these
coordinates are also normal coordinates of the metric $q_{ab}$. 
The motivation
for Eq. (\ref{A:metric}) is given in the remarks below Theorem
\ref{T:main1}.

Fix a non negative scalar field $\tilde \mu$ and a vector field $j^a$
on $M$ with $\mbox{supp}(j^a) \subset \mbox{supp}(\tilde \mu) =
\overline \Omega$, where $\Omega$ is some open set with compact
closure $\overline \Omega \subset \tilde M$, such that its boundary
$\partial \Omega$ is a smooth submanifold of codimension one.
Introduce on $\tilde M$ the fields $\theta$ and $p^{ab}$, with
$p_a{}^a =0$, solutions of
\begin{align}
\label{eq:constr-m}
D_a  p^{ab}  &= - \kappa  j^b, \\ 
\label{eq:constr-h}
L_q (\theta) &= - \frac{p_{ab} p^{ab}}{8\theta^7} 
- \frac{\kappa}{4} \tilde \mu \theta^5,
\end{align}
where $L_q (\theta) := q^{ab} D_aD_b \theta - R\theta/8$, and $D_a$ is
the Levi-Civita connection associated to $q_{ab}$. Indices of
``non-tilde'' tensors are raised and lowered with $q^{ab}$ and $q_{ab}$,
respectively, where $q_{ac} q^{cb} = \delta_a{}^b$.  Fix the boundary
condition
\begin{align}
\label{eq:bc-m}
p^{ij} &= O(r^{-4}),\\
\label{eq:bc-h}
\lim_{r \to 0}r \theta &= 1.
\end{align}

The main part of this work is to prove existence of solution to Eqs.
(\ref{eq:constr-m})-(\ref{eq:bc-h}), and then to show that if the
source function $\tilde \mu$ and $j^a$ are $\Omega$-piecewise and
tangentially smooth, then so are the solutions $p^{ab}$ and $\theta$.
Once these fields $\theta$ and $p^{ab}$ are known the initial data
set is given by the following conformal rescaling
\begin{equation}
\label{eq:conf-resc}
\tilde q_{ab} = \theta^4 q_{ab}, \quad
\tilde p^{ab} = \theta^{-10} p^{ab}, \quad
\tilde j^a = \theta^{-10} j^a.
\end{equation}
Notice that we do not rescale the energy density $\tilde \mu$, but we
do rescale the momentum density $\tilde j^a$. In this way we achieve
both that $\tilde \mu$ be a free data, and that the momentum
constraint decouples the Hamiltonian constraint, respectively. One
can check that if $\theta$ and $p^{ab}$ are solutions of Eqs.
(\ref{eq:constr-m})-(\ref{eq:constr-h}) then the rescaled fields in
Eqs. (\ref{eq:conf-resc}) satisfy Eqs.
(\ref{eq:constr1})-(\ref{eq:constr2}).  One can also check that the
boundary conditions (\ref{eq:bc-m})-(\ref{eq:bc-h}) on $\theta$ and
$p^{ab}$ imply that the rescaled initial data is asymptotically flat.
(See \cite{Beig96,Dain99}.)

Let $\G$ be the Green function of the operator $L_q$ given in Eq.
(\ref{eq:constr-h}), which is defined in Eq.
(\ref{eq:5})-(\ref{eq:6}). Our first main theorem is concerned with
the momentum constraint (\ref{eq:constr-m}), (\ref{eq:bc-m}).
\begin{theorem}
\label{T:main1}
Fix $M$, $\tilde M$, $\Omega$, and $q_{ab}$ as above. Let $s^{ab}$ a symmetric
trace-free tensor in $W^{1,q}(M)\cap C^\infty(\tilde M)$, $q>3$. Let
$\overline p^{ab}$ be given by (\ref{def:pbar}). Assume that 

(i) $\mbox{supp}(j^a) \subset  \overline \Omega
\subset \tilde M$.

(ii) $j^a \in L^q(M)$ and it is $\Omega$-piecewise and
$\Omega$-tangentially smooth.

(iii) Condition (\ref{eq:sycond}) is satisfied.

Then, there exist a unique tensor $p^{ab}$ given by (\ref{eq:singe})
solution of Eq. (\ref{eq:constr-m}), (\ref{eq:bc-m}). Moreover,
$p^{ab}$ is $\Omega$-piecewise and $\Omega$-tangentially smooth and
satisfies
\begin{equation}
  \label{eq:10}
 p_{ab}p^{ab}/\G^7 \in L^2(M). 
\end{equation}
\end{theorem}

The existence part of this theorem is essentially the standard York
splitting (cf. \cite{York73}) adapted to our setting. It is given in
Theorem \ref{T:Mc-exist1}, under a weaker hypothesis. $s^{ab}$ is free
data related to the arbitrary amount of gravitational radiation
that can be added to the system keeping the matter sources fixed.
$\overline p^{ab}$ contains the linear and angular momentum of the
initial data, it can also be prescribed freely unless there are
conformal symmetries. In this case it has to satisfies condition
(iii), which is the corresponding Fredholm condition (see the remark
after Theorem \ref{T:Mc-exist1} for a physical interpretation).  The
regularity part of the theorem is proved in Sec. \ref{s-reg}.

We have chosen the unphysical metric, $q_{ab}$, smooth on $M\setminus
\{i\}$. This is a reasonable physical assumption. However to also
impose smoothness at $i$ it is too restrictive. In this case initial
data for stationary space-times are ruled out (see \cite{Dain01b}). 
The differentiability at $i$ of the unphysical metric is related with
decay at infinity of the associated physical metric $\tilde q_{ab}$,
imposing smoothness at $i$ means a restriction in the fall off which
is, in particular, incompatible with the stationary solutions.  In
order to include these data, we have made the assumption
(\ref{A:metric}). Although the functions $h_{ij}$ and $\hbar_{ij}$
are smooth, the metric belongs to $C^{2,\alpha}(B_{\epsilon})$ but it
does not belong to $C^{3}(B_{\epsilon})$. The data for stationary
space-times have precisely this form (see \cite{Dain01b}). In order to
prove the theorem one certainly does not need smoothness of
$h_{ij}$ and $\hbar_{ij}$, only a finite  number of derivatives. But
the important point is that in order to prove the last part, Eq.
(\ref{eq:10}), it is not enough to require that, for example, $q_{ab}
\in C^{2,\alpha}(M)$. Eq. (\ref{eq:10}) is essential in order to
prove our  second theorem. We prove (\ref{eq:10}) in Theorem
\ref{T:Mc-exist2}. 

\begin{proof}
The metric given by Eq. (\ref{A:metric}) satisfies that $q_{ab} \in
W^{4,p}(M)$, $p > 3/2$. By assumption $j^a$ and $s^{ab}$ belong to
$W^{1,q}(M)$. Therefore Assumption (iii) and Theorem
\ref{T:Mc-exist1} imply that there exists $p^{ab}\in W^{1,q'}(M)$,
$1<q'<3/2$, given by Eq. (\ref{eq:singe}) which solves Eqs.
(\ref{eq:constr-m}), (\ref{eq:bc-m}). The hypothesis on the metric
given in (\ref{A:metric}) and Theorem \ref{T:Mc-exist2} imply Eq.
(\ref{eq:10}). Assumptions (i), (ii), and Theorems \ref{T:p-smooth}
and \ref{T:t-smooth} imply that $p^{ab}$ is $\Omega$-piecewise smooth
and $\Omega$-tangentially smooth.
\end{proof}

In order to write the next theorem we need to define some constants.
Set $C_p= ||p_{ab}p^{ab}/(8\G^7)||_{L^2(M)}$, $\G_{+} =
\max_{\overline \Omega} \G$, $\G_{-} =\min_{\overline \Omega}(\G)$ and
$j_+=\max_{\overline \Omega} \sqrt{j_aj^a}$ . Let $K$, $k$, be the
positive constants defined in Sec. \ref{s-exist}.  They essentially
depend on the metric $q_{ab}$ and the manifold $M$.  Finally, let
$\epsilon_0>0$ solution of the following equation
\begin{equation}
  \label{eq:8}
 \epsilon_0 j_+
 \G_{-}^{-8}=\frac{K}{|\Omega|^{1/2}(\G_{+}+k\epsilon_0^2C_p)^4}, 
\end{equation}
where $|\Omega|$ is the volume of  $\Omega$ with respect to the
metric  $q_{ab}$. There always exists a unique positive solution to
Eq. (\ref{eq:8}), since for $\epsilon_0>0$ the right-hand side of
(\ref{eq:8}) is a positive, decreasing function of $\epsilon_0$ which
goes to zero at infinity.

\begin{theorem} \label{T:main2}
Assume that the hypothesis of Theorem \ref{T:main1} holds.  Let
$p^{ab}$ be the tensor field given in that Theorem. Assume that
$R>0$. Fix a smooth liquid-type state function, $p(\rho)$, with
zero-pressure energy density, $\rho_0>0$, compatible with condition
(iii). Assume that $0 < \partial p /\partial \rho < 1$.  Let $\tilde
\mu$ be such that

(i) $\mbox{supp}(\tilde \mu) = \overline \Omega$. $\tilde \mu
|_{\partial \Omega} = \rho_0$, and $j^a|_{\partial \Omega} =0$.

(ii) $\tilde \mu$ is  $\Omega$-piecewise and
$\Omega$-tangentially smooth.

(iii) For $\epsilon < \epsilon_0$, $\tilde \mu$ satisfies 
\begin{equation}
  \label{eq:9}
 \epsilon j_+
 \G_{-}^{-8}<\rho_0\leq\tilde \mu \leq 
 \frac{K}{|\Omega|^{1/2}(\G_{+}+k\epsilon^2C_p)^4}.
\end{equation}
Then there exist a positive solution $\theta \in C^{1,\alpha}(\tilde
M)$ of Eqs. (\ref{eq:constr-h}), (\ref{eq:bc-h}) with sources given
by $\tilde \mu$ and $\epsilon p^{ab}$, where $0 \leq \epsilon
<\epsilon_0$ and $0 < \alpha < 1$. 

Moreover, the initial data computed with $\tilde q_{ab} = \theta^4
q_{ab}$, $\tilde p^{ab} = \theta^{-10}\epsilon p^{ab}$, $\tilde j^a=
\theta^{-10} \epsilon j^a$, and $\tilde \mu$ is of liquid-type, as
stated in Definition \ref{d-lt-fluid}. They are $\Omega$-piecewise
smooth, and $\Omega$-tangentially smooth, $\tilde q_{ab} \in
C^{1,\alpha}(\tilde M)$, $\tilde p^{ab}\in C^{\alpha}(\tilde M)$. The
fluid 3-velocity, $\tilde v^a$, vanishes at $\partial \Omega$.
\end{theorem}

We use a non-typical conformal rescaling given in Eq.
(\ref{eq:conf-resc}). The positive outcome is that, in this way,
$\tilde \mu$ is essentially free data, and so, we can choose it
constant at $\partial \Omega$. A negative outcome is that this $\tilde
\mu$ must satisfy the bound (\ref{eq:9}).  The upper bound in
(\ref{eq:9}) is related with the existence of solution, given in
Theorem \ref{T:Hc-exist1}.  In the Appendix we give arguments to show
that this bound is only technical, that is, there exist solutions
which do not satisfy it. However, the example presented there suggest
that this bound will be satisfied for every realistic star.  The lower
bound in (\ref{eq:9}) is related to the energy condition. It is
sufficient condition for the dominant energy condition to hold, see
Sec.  \ref{EC}.

A second negative outcome is that, in order to satisfy the liquid-type
constraint (\ref{extra-constr}), we impose $j^a|_{\partial \Omega}=0$,
this implies $\tilde v^a|_{\partial \Omega}=0$. In order to understand
the implications of this condition on the motion of the fluid, assume
that we have a simple, liquid type, fluid solution of Einstein-Euler's
equation. That is, a 4-dimensional Lorentzian metric $g_{ab}$ and a
unit time like vector field $u^a$, representing the fluid 4-velocity,
solutions of Einstein-Euler's equation. The boundary $\cal{B}$ of the
fluid is the 3-dimensional, time like, hypersurface where $p=0$. Since
we have a simple fluid, this implies that $\rho$ is constant on
$\cal{B}$, hence the vector defined by $N^a=g^{ab}\nabla_b \rho$ is
normal to $\cal{B}$, where $\nabla_b$ is the covariant derivative with
respect to $g_{ab}$. By assumption $N^a$ is not zero on $\cal{B}$. Fix
an arbitrary space like foliation, with normal vector $n^a$; let
$\tilde M$ a member of this foliation. Define $\partial \Omega =
\tilde M \cap \cal{B}$, we will assume that both $\cal{B}$ and
$\partial \Omega$ are smooth submanifolds. The 3-velocity $\tilde
v^a$, defined by $u^a=(n^a+\tilde v^a)/\sqrt{1-\tilde v^2}$, will
vanish at $\partial \Omega$ if and only if the following Eqs. hold
\begin{equation}
  \label{eq:17}
  N_an^a|_{\partial \Omega}=0,
\end{equation}
\begin{equation}
  \label{eq:18}
  N_a\omega^a|_{\partial \Omega}=0,
\end{equation}
where $\omega^a=\epsilon^{abcd}u_b \nabla_c u_d$ is the twist of $u^a$
($\epsilon_{abcd}$ is the volume element of $g_{ab}$ and the indexes
are moved with $g_{ab}$). Eq. (\ref{eq:17}) is a condition on the
foliation: the slice $\tilde M$ has to be tangent to $N^a$. Eq.
(\ref{eq:18}) is a condition on $u^a$, independent of the foliation:
the normal component, with respect to the fluid boundary, of the twist
of $u^a$ must vanish on $\partial \Omega$.  Eq. (\ref{eq:18}) is a
consequence of the Frobenius's theorem (see for example \cite{Wald84})
and the fact that $\partial \Omega$ is a smooth submanifold and $N^a$
is hypersurface orthogonal. Note that $\omega^a$ itself can be
different from zero at $\partial \Omega$. Condition (\ref{eq:18}) is
not time-propagated by $u^a$.  This condition is imposed
only on the initial slice, not in the subsequent evolution. Although
it is a restriction, it is not clear if it is a strong restriction or
not.

Another outcome of this particular conformal rescaling is a lack of
uniqueness of solutions to (\ref{eq:constr1})-(\ref{eq:fluid2}) in
terms of the free data.

We do not require that $\Omega$ be connected. A non-connected domain
can describe several compact bodies. 

This is not the most general result one can obtain with these
methods. One can also find solutions which are not piecewise smooth,
but with some finite differentiability in the interior of the support
of $\rho$. One can even obtain solutions where the support of $\rho$
itself has some finite differentiability. The obtainment of such more
general data from the techniques used to get our result does not
present a substantial difficulty but only a greater level of
technical complication, which could obscure the main ideas necessary
to find these type of data. 

\begin{proof}
The upper bound on $\tilde \mu$ given by Eq. (\ref{eq:9})  and
Theorem \ref{T:Hc-exist1} imply that there exists a strictly positive
solution $\theta = \G+ \vartheta \in C^{1,\alpha}(\tilde M)$ of Eqs.
(\ref{eq:constr-h}), (\ref{eq:bc-h}). Hypothesis (i), (ii) and
Theorems \ref{T:p-smooth} and \ref{T:t-smooth} imply that $\theta$ is
$\Omega$-piecewise smooth and $\Omega$-tangentially smooth.

Let $\tilde q_{ab}$, $\tilde p^{ab}$, $\tilde j^a$ be as stated in
Theorem \ref{T:main2}. Those fields are also $\Omega$-piecewise
smooth and $\Omega$-tangentially smooth, and they satisfy $\tilde
q_{ab} \in C^{1,\alpha}(\tilde M)$, $\tilde p^{ab}\in
C^{\alpha}(\tilde M)$. The lower bound in Eq. (\ref{eq:9}) and Lemma
\ref{L:ec} imply that the dominant energy condition is satisfied,
that is $\tilde j_a\tilde j^a < \tilde \mu^2$.  Assumption (i)
implies that the liquid-type constraint (\ref{extra-constr}) is
trivially satisfied. Finally, Theorem \ref{T:inv} implies that Eqs.
(\ref{eq:fluid1})-(\ref{eq:fluid2}) are invertible. The state
function $p(\rho)$ is a smooth function of $\rho$, so Eqs.
(\ref{eq:fluid1})-(\ref{eq:fluid2}) imply that the fields $\tilde
v^a$ and $\rho$ are $\Omega$-piecewise smooth and
$\Omega$-tangentially smooth. Eq. (\ref{eq:fluid2}) and Assumption
(i) imply that $\tilde v^a |_{\partial \Omega} =0$.
\end{proof}

\section{Existence and regularity}
\label{s-er}

\subsection{Hamiltonian constraint}
\label{s-exist}

Consider Eqs. (\ref{eq:constr-h}), (\ref{eq:bc-h}). To obtain a
solution $\theta$ we first transform this problem on $\tilde M$
with a singular boundary condition at $i\in M$, into a regular
problem on $M$ for another function. The metric $q_{ab}$ has strictly
positive scalar of curvature $R$ and the assumption given in Eq.
(\ref{A:metric}) implies that $q_{ab}\in W^{4,p}(M)$, $p>3/2$.
Therefore, Lemma 3.2 and Corollary 3.3 in \cite{Dain99}, imply that
there exist a unique, positive solution $\G \in C^{1,\alpha}(\tilde
M)$ of the equation
\begin{equation}
  \label{eq:5}
 L_q(\G) =-4\pi \delta_i, 
\end{equation}
where $\delta_i$ is  Dirac's delta distribution with support at
$i$. It is also true that  $\G^{-1} \in C^{\alpha}(M)$ and
\begin{equation}
  \label{eq:6}
 \lim_{r\to 0}r\G=1.
\end{equation}
We introduce the function $\vartheta = \theta -\G$. Then, Eq.
(\ref{eq:constr-h}) for $\theta$ on $\tilde M$ becomes the following
equation for $\vartheta$ on $M$,
\begin{equation}
\label{eq:vtheta}
L_q(\vartheta) =-\frac{p_{ab}p^{ab}}{8(\G+\vartheta)^7} 
-\frac{\kappa}{4} \tilde \mu (\G+\vartheta)^{5}.
\end{equation}

Before stating the Theorem concerning existence of solutions to Eq.
(\ref{eq:vtheta}), we need some notation. Given any function $g \in
W^{2,2}(M)$ and the operator $L_q$, we introduce $k$ to be the
constant such that $|g|_{C^0(M)} \leq k\|L_q(g)\|_{L^2(M)}$.  This
constant can be written as $k=c_sc_L$, where the Sobolev coefficient
$c_s$ is the constant such that $|g|_{C^0(M)} \leq c_s
\,\|g\|_{W^{2,2}(M)}$, while $c_L$ is the constant of the elliptic
estimate $\| g\|_{W^{2,2}(M)} \leq c_L\, \|L_q(g)\|_{L^2(M)}$. (See
\cite{Adams,Aubin82}.) We introduce, as well, the constants $ C_p :=
\|p_{ab}p^{ab}/(8\G^7)\|_{L^2(M)}$, and $\G_{+} := \sup_{\overline
\Omega} \G$. Therefore,
\begin{equation}
  \label{eq:7}
p_{ab}p^{ab}/\G^7 \in L^2(M)  
\end{equation}
is equivalent to the condition $C_p < \infty$.

\begin{theorem}{\bf(Existence)}
\label{T:Hc-exist1}
Let $M$ and $\tilde M$ be as in Sec. \ref{S:MR}. Let $q_{ab}$ be a
Riemannian metric on $M$, such that $q_{ab}\in W^{4,p}(M)$, $p>3/2$, 
and $R>0$.  Assume that $p^{ab}$ satisfies that $C_p < \infty$.  Let
$\tilde \mu$ be a positive function of compact support in $\overline
\Omega \subset \tilde M$, such that
\begin{equation}
\label{eq:b-vtheta}
\|\tilde \mu\|_{L^2(\Omega)} \leq \frac{K}{(\G_{+} +k C_p )^4},
\end{equation}
where $K =4^5/(5^5\kappa k)$. Then there exists a non-negative
solution $\vartheta \in W^{2,2}(M)$ of equation (\ref{eq:vtheta}). The
solution is strictly positive unless both $p^{ab}$ and $\tilde\mu$
are zero. Moreover, it satisfies $\vartheta \leq (\G_+ +5k C_p)/4$.
\end{theorem}
\noindent
{\em Remark:} The proof is based on Schauder's fixed-point Theorem
(see for example \cite{zeidler93}): {\em Let $B\subset X$ be a
nonempty, closed, convex set in a Banach space $X$, and $F:B \to B$
be a continuous mapping. If $F(B)$ is precompact, then $F$ has a fix
point}. The construction of the functional $F$ is similar to the one
made in \cite{Dain99} for Theorem 3.4. The only difference is the
choice of the set $B$, and the main work is to prove that for this
choice we have $F(B) \subset B$.
\begin{proof}
Consider $X=C^0(M)$, which is a Banach space under the supremum norm.
Given a constant $c>0$, define $B_c = \{u\in X:\, 0\leq u \leq c
\}$. One can check that $B_c$ is a convex and closed. Define a
non-linear operator $F: B_c \to X$, by setting
\[
F:=L_q^{-1}\circ f
\]
where the $f:B_c \to L^2(M)$ is the continuous map  given by
\begin{equation} \label{eq:fexis}
f(u):=-\frac{p_{ab}p^{ab}}{8(\G+u)^7} -\frac{\kappa}{4} 
\tilde \mu (\G+u)^{5}.
\end{equation}
Under the assumptions $q_{ab}\in W^{4,p}(M)$, $p>3/2$, and $R>0$ it
has been proved in \cite{Dain99} that the non-linear map $F$ is
continuous and $F(B_c)$ is precompact. The only difference between
the map $F$ and the analogous map $T$ defined in \cite{Dain99} is the
second term in the right hand side of (\ref{eq:fexis}). This term is
continuous. Note that $\G$ is singular at $i$ but we assume that
$\tilde \mu$ has support in $\overline \Omega$ and the point  $i$ is
\emph{not} included in $\overline \Omega$.  

We only have to choose the constant ``$c$'' such that $F(B_c)
\subset B_c$.  The rest of the proof shows how to find ``$c$''. In
what follows we will use Lemma 3.1 of \cite{Dain99} many times.

Introduce  the functions  $\varphi_c \in L^2(M)$  and $\phi_c \in
W^{2,2}(M)$ as follows:
\begin{align}
\nonumber
\varphi_c &:= - \frac{p_{ab}p^{ab}}{8\G^7}-
\frac{\kappa}{4} \tilde \mu \,(\G+c)^{5} \\
\nonumber
\phi_c &:= L_q^{-1}(\varphi_c).
\end{align}
Then, for all $u \in B_c$ we have that $f(u) -\varphi_c \geq 0$. This
is equivalent to $L_q(F(u) - \phi_c) \geq 0$, and then $F(u) \leq
\phi_c$. We now choose the best constant ``$c$'' such that $\phi_c
\leq c$. This is done as follows. Given the bound
\begin{align}
\nonumber
\phi_c &\leq k || L_q (\phi_c) ||_{L^2(M)}, \\
\label{eq:phic}
&\leq k \left[C_p + \frac{\kappa}{4}(\G_{+}+c)^5
\| \tilde \mu \|_{L^2(\Omega)} \right]
\end{align}
we impose that the right hand side of (\ref{eq:phic}) be less or equal
to $c$. We then obtain
\begin{equation}
\label{eq:mu+}
\|\tilde \mu \|_{L^2(\Omega)} 
\leq \frac{4}{\kappa k}\,\frac{c- kC_p}{(\G_{+} + c)^5}.
\end{equation}
This inequality has to be valid for some $c$, in particular  for its
maximum value given by $c_0=(\G_{+} + 5k C_p)/4$. Eq. (\ref{eq:mu+})
evaluated at $c_0$ gives Eq. (\ref{eq:b-vtheta}). Therefore, choosing
$B=B_{c_0}$, condition (\ref{eq:b-vtheta}) implies $F(B) \subset B$.
Finally, Schauder's fixed-point Theorem implies that $F$ has a fix
point in $B$. This fix point is the solution $\vartheta$.
\end{proof}

We now show that, under slightly stronger assumptions on the source
functions $\tilde \mu$ and $p^{ab}$, the function $\theta$ belongs to
$C^{1,\alpha}(\tilde M)$. (This differentiability is important for
Theorem \ref{T:t-smooth}.) 
\begin{theorem}{\bf [$C^{1,\alpha}(\tilde M)$-regularity]}
\label{T:Hc-exist2}
Assume the hypothesis on Theorem \ref{T:Hc-exist1} hold, and let
$\theta = \G + \vartheta$, with $\vartheta \in W^{2,2}(M)$ solution
of (\ref{eq:vtheta}). In addition, assume that $\tilde \mu \in
L^q(\Omega)$, with $q>3$, and that $p_{ab}p^{ab}/\G^7 \in
L_{\loc}^q(\tilde M)$. 

Then,  $\theta\in W_{\loc}^{2,q}(\tilde M) \subset 
C^{1,\alpha}(\tilde M)$ is a solution of Eqs.
(\ref{eq:constr-h}), (\ref{eq:bc-h}).
\end{theorem}
\begin{proof}
From the hypothesis on $\tilde \mu$ and $p^{ab}$, we have
$f(\vartheta) \in L_{\loc}^q(\tilde M)$. Elliptic regularity implies
$\vartheta \in W^{2,q}(\tilde M)$. (See \cite{Gilbarg}.) Sobolev
embedding and $q>3$ imply $\vartheta \in C^{1,\alpha}(\tilde M)$.
Therefore, $\G\in C^{1,\alpha}(\tilde M)$ imply that $\theta \in
C^{1,\alpha}(\tilde M)$. Eq. (\ref{eq:6}) implies that $\theta$
satisfies the boundary condition (\ref{eq:bc-h}).
\end{proof}

\subsection{Momentum constraint}
\label{SS:MC}

Consider Eqs. (\ref{eq:constr-m}), (\ref{eq:bc-m}). The main idea is,
as in Sec. \ref{s-exist}, to transform these equations on $\tilde M$
with a singular boundary condition into an equation on $M$ for a
regular variable. Solutions of this regular equation can be found by
the transverse, traceless decomposition of symmetric tensors. See
\cite{mBdE69} for a transverse decomposition, and \cite{York73} for a
transverse, traceless decomposition. See also \cite{Choquet80}, and
references therein.

All this procedure is performed, however, not in Eq.
(\ref{eq:constr-m}) itself, but in a properly conformal rescaled
version of that equation. The new rescaled metric is chosen such that
its Ricci tensor vanishes at $i$. (The restriction that the unphysical
metric, $q_{ab}$, have strictly positive Ricci scalar on $M$ is not
needed in this subsection.) The positive outcome of this new rescaling
is that it is not hard to prove that solutions $p^{ab}$ with
non-vanishing total linear momentum are included.

The plan of this subsection is, first, to introduce some notation;
second, to set up the procedure to prove existence of solutions
$p^{ab}$ in a weak sense (Theorem \ref{T:Mc-exist1}); and third, to
prove that, under a slightly stronger assumption on the matter
source, the solution satisfies Eq. (\ref{eq:7}). (Theorem
\ref{T:Mc-exist2}.)

We start with the new conformal rescaling. Let $M$, $\tilde M$,
$q_{ab}$, $x^i$, $r$, and $B_{\epsilon}$ as in Sec. \ref{S:MR}. Let
$\chi$ be a cut function, that is a smooth function with support in
$B_{2\epsilon}$ and such that $\chi =1$ in $B_{\epsilon}$. Fix on $M$
the metric $\hat q_{ab}$ given by  
\begin{equation}
 \label{eq:h00}
\hat q_{ab}=\omega_0^4 \;q_{ab},
\end{equation}
where the conformal factor $\omega_0$ has the form
\begin{equation}
\label{eq:cf}
\omega_0 = e^{\chi f_0}, \quad f_0 = \frac{1}{2}\,x^j x^k\,L_{jk}(i),
\end{equation}
and we have evaluated at $i$ the tensor field
\begin{equation}
\label{eq:Lab}
L_{ab}:= R_{ab} -\frac{1}{4}R q_{ab}, 
\end{equation}
with $R_{ab}$ the Ricci tensor of $q_{ab}$. Therefore, $\hat q_{ab}=
q_{ab}$ on $M\setminus B_{2\epsilon}$, and they differ only on
$B_{2\epsilon}$. One can check that $\hat R_{abc}{}^d(i)=0$, that is
the Riemann tensor of $\hat q_{ab}$ evaluated at $i$
vanishes. (An explicit computation shows $\hat R_{ab}(i)=0$.
Since $\hat q_{ab}$ is a 3-dimensional metric, $\hat R_{abc}{}^d(i)
=0$.) This property implies that in its associated Riemann normal
coordinate system at $i$, $\hat x^j$, the metric $\hat q_{ab}$ has
the form  
\begin{equation}\label{new-resc}
\hat q_{ij} = \delta_{ij} + O(\hat r^3),\quad \hat \Gamma_i\,^j\,_k =
O(\hat r^2),
\end{equation} 
where $\hat r$ is the geodesic distance from $i$ measured by
$\hat q_{ab}$. This is the reason for doing the new conformal
rescaling.

We complete the rescaling introducing the fields $\hat p^{ab}$ and
$\hat j^a$ as
\[
\hat p^{ab} = \omega_0^{-10} p^{ab}, \quad
\hat j^a = \omega_0^{-10} j^a.
\]
Therefore, Eqs. (\ref{eq:constr-m}), (\ref{eq:bc-m}) transform into
\begin{equation}\label{new-resc-m}
\hat D_a\hat p^{ab} = - \kappa \hat j^b, \quad
\hat p^{ij} = O(\hat r^{-4}),
\end{equation}
where $\hat D_a$ is the metric connection associated to $\hat q_{ab}$.
Latin indices on ``hatted'' quantities represent components in
the coordinate system $\hat x^i$.

We now start the procedure to transform Eq. (\ref{new-resc-m}) on
$\tilde M$ with a singular boundary condition, into an equation on
$M$ for a regular variable. The singular behavior at $i$ of a
solution $\theta$ of Eqs. (\ref{eq:constr-h}), (\ref{eq:bc-h}) was
captured by the Green function $\G$. In the case of Eq.
(\ref{new-resc-m}), the role analogous to $\G$ is played by a tensor
$\overline p^{ab}$. The construction of this tensor field, that
follows, is detailed in \cite{Dain99}, Secs. 4.1-4.2, but we briefly
sketch it here. 

Consider the manifold $(M, \hat q_{ab})$. Let $B_{2 \hat \epsilon}
\subset U$ a ball of $\hat q_{ab}$-geodesic radius $2\hat \epsilon$
centered at $i$, and $\hat \chi$ the associated cut function, that
is, a smooth function that vanishes on $M\setminus B_{2\hat
\epsilon}$ and $\hat \chi =1$ in $B_{\hat \epsilon}$. Let $\hat n_a =
\hat D_a \hat r$. Introduce on $B_{2\hat\epsilon}\setminus \{i\}$
the tensor fields\footnote{In the particular case where $\hat
q_{ab}=\delta_{ab}$, that is the flat metric, these tensor fields
arise as appropriate derivatives of the tensor $\G_{a}{}^{b} :=
(7\delta_{a}{}^{b} +\hat n_a \hat n^b)/(8\hat r)$ which satisfies
$(\L_{\delta}\G_b)^a = (-4\pi \delta_i )\delta_b{}^a$. The operator
$\L_{\delta}$ is defined after Eq. (\ref{eq:14}).}\goodbreak
\begin{align}
\phi_{(1)}^{ab} &= \frac{3}{2\hat r^2}\left[ 2 Q^{(a}\hat n^{b)} 
- (\delta^{ab}- \hat n^a\hat n^b) \hat n_cQ^c\right] \label{Ph1},\\
\phi_{(2)}^{ab}&= \frac{A}{\hat r^3}(\delta^{ab}-3\hat n^a\hat n^b)
\label{Ph2},\\
\phi_{(3)}^{ab} &= \frac{6}{\hat r^3} \hat
n^{(a}\epsilon^{b)cd}J_c\hat n_d
\label{Ph3},\\
\phi_{(4)}^{ab} &= -\frac{3}{2\hat r^4}\left[2P^{(a}\hat n^{b)} 
+ (\delta^{ab}-5\hat n^a\hat n^b)\hat n_cP^c\right] \label{Ph4},
\end{align}
where $A$ is constant, and $P^a$, $J^a$, and $Q^a$ are constants in
the coordinate system $\hat x^i$. Here $\hat n^b = \hat
n_a\delta^{ab}$, and in Riemann normal coordinates, $\hat n^j=\hat
x^j/\hat r$. These tensors are transverse and traceless with respect
to the flat metric. Let $\overline p_{(k)}^{ab} := \hat\chi
(\phi_{(k)}^{ab} -\hat q^{ab} \hat q_{cd} \phi_{(k)}^{cd}/3)$.  
Finally, introduce $\overline p^{ab}$ as follows
\begin{equation}
\label{def:pbar}
\overline p^{ab} := \sum_{k=1}^4 \overline p_{(k)}^{ab}.
\end{equation}
By construction, the tensor $\overline p^{ab}$ depends on
10 parameters, is smooth on $\tilde M$, vanishes on $M\setminus
B_{2\hat\epsilon}$, is symmetric and $\hat q_{ab}$-traceless, and
satisfies $\overline p^{ij} = O(\hat r^{-4})$, as $\hat r \to 0$. 
It also satisfies
\begin{equation}
  \label{eq:11}
  \hat D_a \overline p^{ab}=O(\hat r^{-2})\quad \text{ as } \hat r 
  \rightarrow 0.
\end{equation}
The last equation is obtained as follows: write $\hat D_a \overline
p^{ab}$ explicitly, and then note that first, the tensor fields
$\phi_{(k)}^{ab}$ are divergence and trace free with respect to the
flat metric, and second, that in the coordinates $\hat x^k$ the metric
connection coefficients satisfy (\ref{new-resc}).

We finally recall some needed properties of conformal Killing vector
fields. Consequently, this paragraph is applicable to both $q_{ab}$
and $\hat q_{ab}$. We point out the differentiability of the various
fields, for later purposes. Fix a manifold $(M,q_{ab})$, with $q_{ab}
\in W^{4,p}(M)$, with $p>3/2$. A conformal Killing vector field,
$\xi^a$, is defined by $(\l_{q} \xi)^{ab} :=2[D^{(a}\xi^{b)} - q^{ab}
D_c\xi^c/3] =0$, where $\l_q$ is the conformal Killing operator
associated to the metric $q_{ab}$. There are at most ten conformal
Killing vector fields for a 3-dimensional metric. Given a vector field
$\omega^a \in L^{p'}(M)$, with $p' >1$, we say that it is orthogonal
to $\xi^a$ if
\begin{equation}\label{eq:13}
\int_{M} \xi_{a} \omega^a \; dV=0, 
\end{equation}
where the volume element is computed with the unphysical metric
$q_{ab}$. Notice that the differentiability assumption on the metric
implies that $\xi^a \in C^{2,\alpha}(M)$. This, in turn, with the
H\"older inequality, implies that the integral above is well defined.
We also introduce the conformal Killing data at $i$, that is,
\begin{equation}
\label{eq:ckvdata}
k_a = \frac{1}{6}\,D_a D_b \xi^b(i), \quad
S^a=\epsilon^{abc}D_b \xi_c(i), \quad  
q^a=\xi^a(i), \quad  
a =\frac{1}{3} D_a\xi^a(i). 
\end{equation}
Since $M$ is connected, the integrability conditions for conformal Killing
fields (cf. \cite{Yano57}) entail that these ten ``conformal Killing data
at $i$'' determine the field $\xi^a$ uniquely on $M$.

We have the following existence theorem, which is a generalization of
Theorem 16 proved in \cite{Dain99}.  

\begin{theorem}{\bf (Existence)}
\label{T:Mc-exist1}
Let $M$, and $\tilde M$ be as in Sec. \ref{S:MR}. Assume $q_{ab} \in
W^{4,p}(M)$, $p>3/2$.  Let $\overline p^{ab}$ be defined by
(\ref{def:pbar}), and  and $\hat q_{ab}$ as in (\ref{eq:h00}). Let
$s^{ab} \in W^{1,p'}(M)$ be a symmetric traceless tensor, and $j^a
\in L^{p'}(M)$, with $p' >1$.

(i) If the metric $q_{ab}$ admits no conformal Killing vectors on
$M$, then there exists a unique vector field $w^a \in W^{2, q}(M)$,
with $q=p'$ if $p' < 3/2$ and $1 < q < 3/2$ if $p' \geq 3/2$ such
that the tensor field
\begin{equation}
\label{eq:singe}
p^{ab} = \omega_0^{10}\,\left[\overline p^{ab} + s^{ab} 
+ (\l_{\hat q} w)^{ab}\right]
\end{equation}
satisfies Eqs. (\ref{eq:constr-m}), (\ref{eq:bc-m}).

(ii) If the metric $q_{ab}$ admits a conformal Killing vector
$\xi^a$ on $M$, corresponding to the conformal Killing data given in
Eq. (\ref{eq:ckvdata}), then a vector field $w^a$ as specified above
exists if and only if  the following condition holds,
\begin{equation}
\label{eq:sycond}
P^a\,k_a + J_a\,S^a + A\,a + (P^b\,L_b\,^a(i) + Q^a)\,q_a = \kappa
\int_M j_a\xi^a  \;dV,    
\end{equation}
where the constants $P^a$,  $J_a$, $A$, and $Q^a$ characterize the
tensor $\overline p^{ab}$ as in (\ref{Ph1})-(\ref{def:pbar}).    
\end{theorem}

\begin{proof}
Because of (\ref{eq:11}) we can consider $\hat D_a(\overline
p^{ab}+s^{ab})$ as a function in $L^q(M)$, $1<q<3/2$. The equation $D_a
p^{ab}=-\kappa j^b$ is equivalent to 
\begin{equation}
  \label{eq:12}
  \hat D_a [\overline p^{ab} +s^{ab} +(\l_{\hat q} w)^{ab}]
  =-\kappa \omega_0^{-10} j^b,
\end{equation}
which can be written like
\begin{equation}
  \label{eq:14}
({\bf L}_{\hat q} w)^b =  -\kappa\omega_0^{-10} j^b- \hat D_a(\overline
p^{ab}+s^{ab}),
\end{equation}
where $({\bf L}_{\hat q} w)^a := \hat D_b ({\cal L}_{\hat q}w)^{ab}$
is an elliptic operator. Its kernel consists of all conformal Killing
vectors, $\xi^a$, of $\hat q_{ab}$, and so, of $q_{ab}$. Following 
\cite{Dain99}, one can prove that the right hand side of
(\ref{eq:14}) is orthogonal [in the sense given in (\ref{eq:13})] to
every conformal Killing vector field, $\xi^a$, if and only if Eq.
(\ref{eq:sycond}) holds. Therefore, the assumptions on the metric,
$q_{ab}$, and the Fredholm alternative for this operator imply there
exists a unique solution $w^a \in W^{2,q}(M)$. (For smooth metric
this is a standard result, for metric in the Sobolev space
$W^{4,p}(M)$ see \cite{Cantor}.)
\end{proof}

The quantities $P^a$ and $J^a$ in tensor $\overline p^{ab}$ represent
the total linear and angular momentum of the data. These quantities
can be prescribed freely in case (i), so they are not related with the
matter sources $j^a$. The interpretation is that gravitational waves
can carry an arbitrary amount of linear and angular momentum.  In the
case that the unphysical metric has conformal symmetries these
quantities are restricted by condition (\ref{eq:sycond}). In order to
understand this condition, consider the case where only one Killing
vector $\xi^a$ exists, and it is a rotation. That is, only $S^a$ is
different from zero. We can always choose $S^a$ to be a unit vector.
(This vector is parallel to the axis of the rotational symmetry.)
Construct the following initial data: first, choose any $J^a$ pointing
in the same direction as $S^a$, and second, choose the other part of
the free data preserving the symmetry. Then, all the fields in the
initial data set have this symmetry, and therefore the whole
space-time obtained from this initial data set will also have a
Killing vector $\xi^a$, suitably extended outside the initial
hypersurface. Condition (\ref{eq:sycond}) reduces to
\begin{equation}
\label{eq:16} J=\kappa \int_M j_a\xi^a  \;dV, 
\end{equation} 
where $J=\sqrt{J_aJ^a}$. Eq. (\ref{eq:16}) is just the standard Komar
integral. (See for example \cite{Wald84}.) This is consistent with
the interpretation that axially symmetric gravitational waves do not
carry angular momentum.

Notice that, with the assumptions we have made, we do not even know
if $w^a$ is a continuous vector field. We start with the final part
of this subsection, namely, to show that under a slightly stronger
assumption on the differentiability of $j^a$ on $M$, and on the
metric $q_{ab}$ at $i$, the tensor $p^{ab}$ given by (\ref{eq:singe})
satisfies Eq. (\ref{eq:7}).   We have assumed that $q_{ab} \in
W^{4,p}(M)$. We now impose on the metric an extra condition given in
Eq. (\ref{A:metric}). Then, we have the following result:
\begin{theorem}{\bf(Regularity on $\tilde M$)}
\label{T:Mc-exist2}
Assume that hypothesis in Theorem \ref{T:Mc-exist1} hold.  Assume that
the metric satisfies (\ref{A:metric}). If $j^a \in L^q(M)$, $s^{ab}\in
W^{1,q}(M)$, where $q>3$, then, $w^a \in C^{1,\alpha}(\tilde M)$ and
the tensor $p^{ab}$ satisfies $p_{ab}p^{ab} /\G^7 \in L^2(M)$.
\end{theorem} 

That $w^a \in C^{1,\alpha}(\tilde M)$ is deduced from standard
elliptic regularity theorems. The second part is more difficult. The
problem is that $(\l_{\hat q}w)^a$ is not continuous at $i$, and so
conditions that involve products of tensors are difficult to prove.
Since  the origin of the discontinuity in $(\l_{\hat q}w)^a $ is the
singular behavior of  $\overline p^{ab}$, which we  know
explicitly, we proceed as follows. We split $w^a$ into a regular 
part at $i$ (called $\omega^a$ in the proof) plus some divergent
terms. These divergent terms are explicitly computed in terms of
$\overline p^{ab}$ by an integration procedure based on Meyers'
result \cite{Meyers}. Finally we show that this $\omega^a$ satisfies
a linear elliptic equation with a source in $L^q(B_{\epsilon})$ with
$q>3$. Therefore it is $C^{1,\alpha}$ at $i$. Once this splitting
near $i$ on $w^a$ is established, condition (\ref{eq:7}) is proved by
explicit computation.

\begin{proof} 
The source $j^a \in L^q(M)$, while $\overline p^{ab}$, being smooth
on $\tilde M$, belongs to $L^q_{\loc}(\tilde M)$. Therefore, standard
elliptic regularity theorems given in \cite{Gilbarg} imply $w^a \in
W_{\loc}^{2,q}(M) \subset C^{1,\alpha}(\tilde M)$. 

We now begin the proof of the second part of the theorem. We work
always with the rescaled metric $\hat q_{ab}$ and its corresponding
covariant derivative $\hat D_a$. We  explicitly compute the divergent
terms of $w^a$ at $i$. These terms are appropriate Meyers' potentials
of the divergent terms present on $\hat D_i\overline p_{(k)}^{ij}$.
(See Lemma \ref{L:Meyers} below.) 

Let $B_{\hat \epsilon} \subset M$ be an open ball centered at $i$ of
geodesic radius $\hat \epsilon >0$. Let us choose $\hat \epsilon$
small enough such that the metric has the form (\ref{new-resc}) in
Riemann normal coordinates at $i$, and the cut function $\hat \chi$
is identically equal to 1 in $B_{\hat \epsilon}$. An explicit
computation shows 
\[
\hat D_i\overline p^{ij} =
\sum_{k=2}^3 \frac{\mathring{p}^j_{(k)}}{\hat r^{k-1}} 
+ \varphi^j,
\]
with $\varphi^i=O(1)$ and smooth on $B_{\hat \epsilon} \setminus
\{i\}$, and continuous at $i$. The $\mathring p^i_{(k)}$ are
functions of $\hat n_a$. We adopt the convention that a small circle
over a quantity means that this quantity depends smoothly on $\hat
n_a$, and does not depend on $\hat r$.

Let $V_{(k)}^i$ denote the Meyers potentials of
$\mathring{p}^j_{(k)}/\hat r^{(k-1)}$, for each $k=2,3$, that is,
vector fields $V_{(k)}^i=\left(\sum_{l=0}^2 [\ln (\hat r)]^l
\mathring{v}_{(kl)}^i\right)/\hat r^{(k-3)}$  defined on $B_{\hat
\epsilon}$, with $\mathring v^i_{(kl)}$ appropriate functions of
$\hat n_a$ that can be explicitly computed in terms of $\mathring
p^i_{(k)}$, and satisfying
\[
(\L_{\delta}V_{(k)})^i = \frac{\mathring{p}^i_{(k)}}{\hat r^{(k-1)}}.
\]
So, here is our decomposition of the vector field $w^i$, on
$B_{\hat \epsilon}$,
\[
w^i = \sum_{k=2}^3 V_{(k)}^i  + \omega^i.
\]
The rest of the proof shows that $\omega^i$ is indeed differentiable
at $i$.

Thus $\omega^i$ satisfies, 
\[
(\L_{\hat q}\omega)^i = -\kappa \omega_0^{-10}j^i- \sum_{k=2}^3 
(\tilde \L_{\hat q} V_{(k)})^i  -\varphi^i - \hat D_js^{ij},
\]
where $(\tilde \L_{\hat q}V_{(k)})^i := (\L_{\hat q}V_{(k)})^i -
(\L_{\delta}V_{(k)})^i$. One can check that $(\tilde \L_{\hat
q}V_{(k)})^i = O(\hat r^{-(k-3)})$. Therefore the terms $(\tilde
\L_{\hat q}V_{(k)})^i$ with $k=2,3$, belong to $L^q(M)$ with $q > 3$.
Standard elliptic regularity implies that $\omega^i \in W^{2,q}(M)
\subset C^{1,\alpha}(M)$.

We have proved that the solution $w^i \in C^{1,\alpha}(\tilde M)$ has
the following expression in $B_{\hat \epsilon}$,
\[
w^i = \sum_{k=2}^3 \frac{1}{\hat r^{k-3}} \left(\sum_{l=0}^2 
[\ln (\hat r)]^l \mathring{v}_{(kl)}^i\right) + \omega^i.
\]
Notice that the conformal factor $\omega_0$ is smooth on $M$, then an
explicit computation implies that $p^{ab}$ given by Eq.
(\ref{eq:singe}) satisfies Eq. (\ref{eq:7}).
\end{proof}

We present here the generalization of Meyers' result, used in the
proof of Theorem \ref{T:Mc-exist2}. 
\begin{lemma}\label{L:Meyers}
{\bf (Meyers' potential for $\L_{\delta}$)}
Consider the manifold $(\R^3,\delta_{ab})$, with $\delta_{ab}$ the
flat metric, $\partial _a$ the metric connection, and let
$(\L_{\delta} V)^a= \partial_b\partial^bV^a +
\partial^a\partial_bV^b/3$.  Consider the equation
\begin{equation}\label{eq:Meyers}
(\L_{\delta}V)^a = r^{k-2} \sum_{l=0}^{\ell}[\ln (r)]^l\;
\mathring{p}_{(l)}^a(n) 
\end{equation}
where $\ell\geq 0$ is a fix integer, $r$ the geodesic distance from
an arbitrary point $p\in \R^3$, and $\mathring{p}_{(l)}^a(n)$ is a
$C^{K,\alpha}(\R^3)$ function of $n_a = \partial _a r$, with $K\geq
0$.

Then, there exists $C^{K+2,\alpha}(\R^3)$ functions
$\mathring{V}_{(l)}^a(n)$, with $l=0,\cdots,\ell+2$, such that
\begin{equation}\label{def:V}
V^a = r^k \sum_{l=0}^{\ell+2}  \left[\ln(r)\right]^l \;
\mathring{V}_{(l)}^a(n)
\end{equation}
is a solution of (\ref{eq:Meyers}).
\end{lemma}

\begin{proof}
We look for solutions of Eq. (\ref{eq:Meyers}) of the form
\[
V^a = v^a -\frac{1}{4} \partial^a \lambda
\]
with
\begin{align*}
\partial_b\partial^b v^a &= r^{k-2} \sum_{l=0}^{\ell}[\ln (r)]^l\;
\mathring{p}_{(l)}^a(n)\\
\partial_a\partial^a \lambda &= \partial_av^a.
\end{align*}
Lemma 4 in \cite{Meyers} implies that there exist
$C^{K+2,\alpha}(\R^3)$ functions $\mathring v_{(l)}^a$, with
$l=0,\cdots,\ell+1$, such that
\[
v^a = r^k \sum_{l=0}^{\ell+1}  \left[\ln(r)\right]^l \;
\mathring{v}_{(l)}^a(n)
\] 
satisfies the first equation above. Then, one can explicitly compute
$\partial_av^a$, and again Lemma 4 in \cite{Meyers} implies that   
there exist
$C^{K+3,\alpha}(\R^3)$ functions $\mathring
\lambda_{(l)}$, with $l=0,\cdots,\ell+2$, such that
\[
\lambda = r^{k+1} \sum_{l=0}^{\ell+2}  \left[\ln(r)\right]^l \;
\mathring{\lambda}_{(l)}(n)
\]
is a solution of the second equation above. Therefore, an explicit
computation gives (\ref{def:V}).
\end{proof}

\subsection{Local regularity}
\label{s-reg}

Consider a solution $\theta$, $p^{ab}$ of Eqs.
(\ref{eq:constr-m})-(\ref{eq:bc-h}). Assume now that the free data
$s^{ab}$, $\tilde \mu$ and $j^a$ are $\Omega$-piecewise and
tangentially smooth.  In the first part of this subsection we then
prove that the fields $\theta$ and $p^{ab}$ are $\Omega$-piecewise
smooth. (Theorem \ref{T:p-smooth}.) This proof is based on standard
elliptic regularity theorems. In the second part of this subsection
we prove that these fields are also $\Omega$-tangentially smooth.
This result is split into two parts, first for linear elliptic
systems, (Lemma \ref{L:tr}), and then for Eqs.
(\ref{eq:constr-m})-(\ref{eq:bc-h}). (Theorem \ref{T:t-smooth}.)

\begin{theorem}{\bf ($\Omega$-piecewise smooth)}
\label{T:p-smooth} Let both, $q_{ab}$ and $s^{ab}$, in
$C^\infty(\tilde M)$. Let $\theta$ and $p^{ab}$ be solutions of
\eqref{eq:constr-m}-\eqref{eq:bc-h} given by Theorem
\ref{T:Hc-exist2} and Theorem  \ref{T:Mc-exist1}.  If the source
functions $\tilde \mu$ and $j^a$ are $\Omega$-piecewise smooth then
so are $\theta$ and $p^{ab}$.
\end{theorem}

\begin{proof}
By the assumption $q_{ab}\in C^\infty(\tilde M)$ we have that the two
elliptic operators $L_q$ and $\mathbf{L}_q$ have smooth coefficients
in $\tilde M$.  Applying the standard interior elliptic regularity to
the domains $\overline \Omega$ and $\tilde M\setminus \Omega$ we
obtain that if $j^a$ is $\Omega$-piecewise smooth then $w^a$ is also
$\Omega$-piecewise smooth. Because $\overline p^{ab}\in
C^{\infty}(\tilde M)$, and by assumption $s^{ab}\in C^{\infty}(\tilde
M)$, then $p^{ab}$ is $\Omega$-piecewise smooth.

In the case of $\theta$ we note first that by the elliptic regularity
$\G$ is smooth in $\tilde M$. Consider now equation \eqref{eq:vtheta}
for $\vartheta$. Denote by $f(x,\vartheta)$ the right hand side of
this equation. By the assumption on $\tilde \mu$ and the previous
argument regarding $w^a$, we have that the function $f(x,\vartheta)$
satisfies the following property: if $\vartheta$ belongs to
$C^{s,\alpha}(\overline \Omega)$, [or to $C^{s,\alpha}(\tilde M
\setminus \Omega)$] then the composition $f(x,\vartheta(x))$ defines
a function that belongs to $C^{s,\alpha}(\overline \Omega)$ [or to
$C^{s,\alpha}(\tilde M \setminus \Omega)$, respectively]. By  Theorem
\ref{T:Hc-exist2} we know that the solution $\vartheta \in
C^{1,\alpha}(\tilde M)$. (The argument works also with $\vartheta
\in C^{\alpha}(\tilde M)$.) Then we make an iteration, applying the
elliptic regularity for the domains $\overline \Omega$ and $\tilde
M\setminus \Omega$ in each step, to obtain that $\vartheta$ is
$\Omega$-piecewise smooth. Therefore, so is $\theta$. 
\end{proof}

Let $\Omega \subset \Omega'$, and $V^a$ a smooth vector field on
$\Omega'$. Let $u$ any tensor field on $M$.  Denote $V^{(0)}(u) := u$,
$V^{(1)}(u) := V^aD_au$, and $V^{(k)}(u) := V^aD_a[V^{(k-1)}(u)]$, for
$k\geq 1$. In order to prove tangential regularity we prove  first the 
following lemma. 

\begin{lemma} 
\label{L:tr}
Let $L$ be a linear elliptic operator of second order on some open,
bounded set $\Omega' \subset \tilde M$ with smooth coefficients. Let
$V^a$ be a smooth vector field on  $\Omega's$, with $\overline \Omega
\subset \Omega'$. Let $u\in W^{2,p}(\Omega')$, with $p>1$, be a
tensor field on $\Omega'$ solution of the elliptic equation $L(u)=f$.
Let $k\geq 0$, an integer.

(i) If $V^{(k)}(f) \in L^{p}(\Omega')$, then $V^{(k)}(u) \in
W^{2,p}(\Omega')$.

(ii) If $V^{(k)}(f) \in C^{\alpha}(\Omega')$, then $V^{(k)}(u) \in
C^{2,\alpha}(\Omega')$.
\end{lemma}
\begin{proof}
The proof is by induction on $k$. Consider the part (i) of the Lemma.
The case $k=0$ is the standard interior elliptic regularity. See
\cite{Gilbarg} for second order elliptic equations and
\cite{Agmon64,Douglis55,Morrey66} for systems. Assume now that (i) is
true for $k-1$. Consider now the following identity
\begin{equation}
\label{eq:binom}
V^{(k)}(L(u)) = \sum_{l=0}^k  \binom{k}{l} 
{}^{(l)}[V,L](V^{(k-l)}(u)),
\end{equation}
where $\binom{k}{l} =k!/[l!(k-l)!]$ and we have introduced the
notation 
\begin{align}
{}^{(0)}[V,L](u) &:= L(u) \nonumber \\
{}^{(1)}[V,L](u) &:= [V,L](u) = V(L(u))-L(V(u)) \nonumber \\
{}^{(l+1)}[V,L](u) &:= [V,{}^{(l)}[V,L]](u). \nonumber
\end{align}
Notice that, for all $l\geq 0$, the operator ${}^{(l)}[V,L]$ is a
second order operator with smooth coefficients on $\tilde M$. Assume
now that $V^{(k)}(f) \in L^{p}(\Omega')$, and $V^{(l)}(u) \in
W^{2,p}(\Omega')$, for all $0 \leq l\leq k-1$. If we write the
identity (\ref{eq:binom}) as
\begin{equation}
L(V^{(k)}(u)) = V^{(k)}(f) - \sum_{l=1}^k \binom{k}{l} 
{}^{(l)}[V,L](V^{(k-l)}(u))
\end{equation}
then all the terms in the right hand side belongs to $L^p(\Omega')$.
Then the elliptic regularity theorems imply that $V^{(k)}(u)\in
W^{2,p}(\Omega')$, $p>1$. The case (ii) is similar.
\end{proof}

\begin{theorem}{\bf ($\Omega$-tangentially smooth)}
\label{T:t-smooth}
Assume the hypothesis on Theorem \ref{T:p-smooth}. If $\tilde \mu$
and $j^a$ are $\Omega$-tangentially smooth then so are the fields
$\theta$ and $p^{ab}$ which solve Eqs.
\eqref{eq:constr-m}-\eqref{eq:bc-h}.
\end{theorem}

\begin{proof}
Fix $\Omega'$, any open set in $\tilde M$ such that $\overline \Omega
\subset \Omega'$. Let $V^a$ to be the tangent vector field
$V^a_{\partial \Omega}$ defined in Sec. \ref{liquidtype}.   
Since Eq. (\ref{eq:14}) is linear, Lemma
\ref{L:tr} implies that $w^a$ is $\Omega$-tangentially smooth.

Eq. (\ref{eq:vtheta}) is semi-linear. However, there exists a
solution $\vartheta \in W^{2,q}(\Omega')$ for $q>3$. Therefore,
$\vartheta \in C^{1,\alpha}(\Omega ')$. This is the subtle step.
Because $\vartheta \in C^{1,\alpha}(\Omega')$, it implies that
$V(f(x,\vartheta(x))) \in L^q(\Omega')$, for $q>3$. The reason is
that when we compute $V(f)$, appear terms of the form ``function in
$L^{p}(\Omega')$'' times ``$V(\vartheta)$.'' If $\vartheta$ was only
continuous, then, these terms would be not, in general, in
$L^p(\Omega')$. Then Lemma \ref{L:tr} implies that $V(\vartheta) \in
W^{2,q}(\Omega')\subset C^{1,\alpha}(\Omega')$. Thus
$V^{(2)}(\vartheta) \in C^{\alpha}(\Omega')$. Then $
V^{(2)}(f(x,\vartheta(x))) \in L^q(\Omega')$ and we obtain
$V^{(2)}(\vartheta) \in C^{1,\alpha}(\Omega')$. Iterating this
argument, the conclusion follows.
\end{proof}

\section{Further requirements}
\label{S:FR}

\subsection{Energy condition}
\label{EC}

In order to understand the origin of this discussion on energy
conditions it is useful to compare the usual procedure to find
solutions of the constraint equations with matter sources.  In that
procedure, one rescales both, the energy density $\tilde \mu$ as well
the momentum current density $\tilde j^a$. The rescaled $\tilde j^a =
\theta^{-10} j^a$ is fixed from the requirement that the momentum
constraint be independent of $\theta$, while the rescaled $\tilde \mu
= \theta^{-8} \mu$ is chosen such that $\tilde j/ \tilde \mu = j /
\mu$, where we have introduced $\tilde j := \sqrt{\tilde q_{ab}\tilde
j^a \tilde j^b}$, and  $j := \sqrt{q_{ab} j^a j^b}$. Therefore, if
the energy condition is satisfied by the rescaled fields $j^a$ and
$\mu$, with respect to the unphysical metric, then the physical
fields $\tilde j^a$ and $\tilde \mu$ do satisfy the energy condition.
That is the usual procedure. Here we can not rescale the energy
density, because we need the extra condition that the physical energy
density $\tilde \mu$ be constant at the border of its support. 

We show here that the lower bound for $\tilde \mu$ given in  (iii) in
Theorem \ref{T:main2} is sufficient to guarantee that the physical
matter fields satisfy the dominant energy condition. The idea is
this: Initial data with $j^a=0$ and $\tilde \mu$  satisfy trivially
the energy condition. Therefore the same applies for arbitrary small
$j$. Condition (iii)  is just a rough bound on the smallness on $j$
that also guarantees the energy condition. Let $\G_{-} =
\min_{\overline \Omega} (\G)$, with $\G$ the Green function solution
of Eqs. (\ref{eq:5})-(\ref{eq:6}). Then we have the following:
\begin{lemma}\label{L:ec}
Let $M$, $\tilde M$, $q_{ab}$, $\tilde \mu$, and $p^{ab}$ be as in
Theorem  \ref{T:Hc-exist2}. Let $\theta$ be the corresponding
solution of Eqs. (\ref{eq:constr-h}), (\ref{eq:bc-h}). Let $\tilde
\mu$ and $j^a$ have support in $\Omega$ and in  $C^0(\overline
\Omega)$.  If $j < \rho_0 (\G_{-})^8$ then the fields $\tilde q_{ab}
= \theta^4 q_{ab}$, $\tilde j^a=\theta^{-10}j^a$, and $\tilde \mu$,
satisfy $\tilde j < \tilde \mu$. 
\end{lemma}
\begin{proof}
Since $\vartheta$  is positive (see Theorem \ref{T:Hc-exist1}),
$\theta \geq \G$ on $\tilde M$. Then we have
\[
\tilde j = \theta^{-8} j \leq (\G_{-})^{-8} j 
<  \rho_0  \leq \tilde \mu.
\]
\end{proof}

\subsection{Inversion of Eqs.
(\ref{eq:fluid1})-(\ref{eq:fluid2})}  

\label{s-Bw}

We show here that Eqs. (\ref{eq:fluid1})-(\ref{eq:fluid2}) are
invertible, that is given the functions $\tilde \mu$, $\tilde j^a$
then there exists unique functions $\rho$, $\tilde v^a$ satisfying
these equations. In other words, the fluid 4-momentum density as seen
by an arbitrary observer determines the fluid co-moving 4-momentum
density. It turns out that the proof is not obvious and we did not
find it in the literature.

The main difficulty is that the map defined by Eqs.
(\ref{eq:fluid1})-(\ref{eq:fluid2}) is non-linear. Furthermore, it
contains an unknown function, the state function, subject to
minimally restrictive properties. Since $\tilde v^a$ and $\tilde
j^a$ are parallel, these equations reduce to
\begin{align}
\label{eq:mu}
\tilde \mu &= \frac{\rho + p \tilde v^2}{1-\tilde v^2} \\
\label{eq:j}
\tilde j &= \frac{(\rho+p)\tilde v}{1-\tilde v^2},
\end{align}
with $\tilde j = \sqrt{\tilde j_a\tilde j^a}$ and $\tilde v =
\sqrt{\tilde v_a\tilde v^a}$. We define the map $\Phi$ between
subsets of $\mathbb{R}^2$ as
\begin{equation}
  \label{eq:1}
  \Phi(\rho, \tilde v)=\left (\frac{\rho + p \tilde v^2}{1-\tilde
      v^2}, \frac{(\rho+p)\tilde v}{1-\tilde v^2} \right).
\end{equation}
Eqs. (\ref{eq:mu})-(\ref{eq:j}) can be rewritten as $(\tilde \mu,
\tilde j) = \Phi(\rho, \tilde v)$. Given a positive constant $\rho_0$
we define the following two subsets of $\mathbb{R}^2$
\begin{align}
  \label{eq:2}
  D :&=\{(\rho,\tilde v) \in \R^2: \rho_0\leq \rho, \quad 0 \leq
  \tilde v <1\} \\
  I:&=\{ (\tilde \mu,\tilde j)\in \R^2: \tilde \mu_0(\tilde j) 
  \leq \tilde \mu, \quad 0\leq \tilde j\},
\end{align}
where $\tilde \mu_0(\tilde j) := \rho_0/2 + \sqrt{\rho_0^2/4+\tilde
j^2}$.  Our result is:

\begin{theorem}
\label{T:inv}
Let $p(\rho)$ be a $C^{1}$ state function such that: (i) $p(\rho)
\geq 0$ for $\rho \geq \rho_0 >0$; (ii) $p(\rho_0)=0$; (iii) $0<
\partial p/\partial \rho <1$. Then, the map $\Phi:D\to I$ is a
diffeomorphism.
\end{theorem}

\begin{proof} 

First we prove that $\Phi$ is bijective.

{\em Surjectivity:} 
The essential tool is Brouwer's fixed-point theorem. (See for
example \cite{zeidler93}.) 
Eqs. (\ref{eq:mu})-(\ref{eq:j}) are equivalent to
\begin{align}
\label{eq:rho}
\rho &= \tilde \mu - \tilde j \tilde v \\
\label{eq:v}
\tilde v &= \frac{\tilde j}{\tilde \mu} \, 
\frac{\rho + p \tilde v^2}{\rho+p}.
\end{align}
Fix a point $(\tilde \mu,\tilde j) \in I$. Consider the map $F$ given
by 
\[
F(x,y) := \left[(\tilde \mu -\tilde j y), 
\left(\frac{\tilde j}{\tilde \mu} \, \frac{x + p(x) y^2}{x+p(x)}
\right)\right]. 
\] 
Introduce the compact convex  set $C:=[\rho_0,\tilde \mu]\times [0,1]
\subset \R^2$.

We claim that $F:C \to C$. We write  $F(x,y) = (F_1(x,y), F_2(x,y))$.
Then, by definition of $I$,  $\tilde j/ \tilde \mu < 1$ and so, for
all $(x,y) \in D$, we have $0\leq F_2(x,y) < 1$. We now show that,
for all $(x,y) \in D$,   $\rho_0 \leq F_1(x,y) \leq \tilde \mu$. The
assumption $0 \leq y \leq 1$  implies $\tilde \mu - \tilde j \leq
\tilde \mu-\tilde j y \leq \tilde \mu$. But  $\rho_0 \leq \tilde
\mu_0(\tilde j) - \tilde j \leq \tilde \mu - \tilde j$. Therefore 
$\rho_0 \leq \tilde \mu -\tilde j y \leq \tilde \mu$.

The map $F$ is also continuous. Therefore, by Brouwer's fixed-point
theorem, there exists a fix point $F(x,y)= (x,y)$. 

Notice that $\tilde j < \tilde \mu_0(\tilde j) \leq \tilde \mu$
implies that there exists no fix point of the form $(x,1)$.
Therefore, we conclude that, given a point $(\tilde \mu,\tilde j)\in
I$, there exists a point $(\rho,\tilde v)\in D$ which solves Eqs.
(\ref{eq:mu})-(\ref{eq:j}).

{\em Injectivity:} Consider Eqs. (\ref{eq:mu})-(\ref{eq:j}), written
as
\begin{align*}
\tilde \mu &= \rho + \tilde j \tilde v \\
\tilde j &= \frac{(\rho+ p)\tilde v}{1- \tilde v^2}.
\end{align*}
Assume that there exist two points $(\rho_1,\tilde v_1)$ and
$(\rho_2,\tilde v_2)$ which are solutions of these above equations
for the same value of $(\tilde \mu, \tilde j)$. If $\tilde v_1 =0$
then the second equation above implies $\tilde j=0$, and so $\tilde
v_2 =0$, which in turn implies $\rho_1=\rho_2$. If $\tilde v_1=\tilde
v_2$ then the first equation below implies $\rho_1=\rho_2$.

Assume now that $\tilde v_1 \neq 0$, $\tilde v_2 \neq 0$, and $\tilde
v_1 \neq \tilde v_2$. Then
\begin{align*}
(\rho_2-\rho_1) + \tilde j(\tilde v_2 -\tilde v_1) &=0 \\
(\rho_2-\rho_1)(1-\nu^2) &= \tilde j \left[\frac{1-(\tilde
v_2)^2}{\tilde v_2} - \frac{1-(\tilde v_1)^2}{\tilde v_1}\right],
\end{align*}
where $\nu^2 = (\partial p/\partial \rho)|_{\rho'}$, with  $\rho' \in
[\rho_1,\rho_2]$ and we have used the mean value theorem for
$p(\rho)$. Then the above equations and the  assumptions on $\tilde
v_1$ and $\tilde v_2$ imply
\[
\nu^2 \tilde v_1\tilde v_2 =1.
\]
But by assumption $\nu^2 <1$, so that we have a contradiction.
Therefore, injectivity follows.

It remains to prove that $\Phi$ and $\Phi^{-1}$ are differentiable.
By direct computation and the assumptions on $p(\rho)$ one can check
that the derivative map of  $\Phi$ is invertible at each point of
$D$. Then, by  the inverse function theorem, $\Phi^{-1}$ is also
differentiable. 
\end{proof} 

Notice that the proof fails if $\rho_0=0$ because the derivative of
$\Phi$ is not invertible at this point. This will be the case for an
equation of state of the form $p=a\rho^\gamma$, where $a$ and
$\gamma$  are constants. On the other hand, in this work we are
interested in equations of state of liquid-type, i.e. such that the
pressure vanish for a positive value of the density at the border of
the fluid. For example $p=a[(\rho/\rho_0)^\gamma-1]$. For suitable
constants  $a$ and $\rho_0$ this equation describes water. (See
\cite{courant76}.)

\section{Discussion}
\label{s-d}

The principal interest in the initial data given here is to use them
to set up an initial value formulation. This formulation should be
able to describe isolated, nearly static fluid bodies. That is why we
have concentrated on finding liquid-type data and, inside this class,
the smoothest possible data, i.e. the simplest to evolve. We have
shown here that these data are not simple to obtain.

The discontinuity of the fluid energy density at the boundary of its
support and extra constraints at that boundary [see Eq.
(\ref{extra-constr})] were the main difficulties. One main idea was
to not rescale the fluid energy density and so, being  free data,
trivially solve the extra constraint Eq. (\ref{extra-constr}). (While
also requires that the fluid 3-velocity vanish at the body boundary.)
This unconventional rescaling of the fluid fields introduces
difficulties in the task of finding solutions to the Hamiltonian
constraint. These difficulties were solved in Theorem
\ref{T:Hc-exist1}. Smoothest liquid-type data are almost-smooth, i.e.
smooth except in the normal direction to the body boundary. The main
step in establishing this result is Lemma \ref{L:tr}. The rest is
standard elliptic regularity.

We have shown that at the body boundary, the first fundamental form
is only in $W^{2,q}_{\loc}(\tilde M)$, $q>3$. This differentiability
is below the threshold required by the known existence theorems on
symmetric hyperbolic equations to prove existence of solutions
associated with such a data. Theorems in Sec. 5.1 in
\cite{meT91} require initial data in $W^{s,2}(\Omega')$, with $s >
5/2$, where $\Omega' \subset \R^3$, is open and bounded. (See
\cite{sK01} for a related improvement of this result and also
\cite{sK01b} for a discussion on the possible future development of
the subject) Imbedding Theorems imply that $W^{s,2}(\Omega')
\subset W^{2,p}(\Omega')$, with $p>3$, but not in the other way
around. Therefore, data in $W^{2,p}(\Omega')$ is not enough for the
known theorems to guarantee existence of solutions. 

We guess two possible ways to set up an initial value formulation for
these liquid-type bodies. The first one is to study in detail the
Einstein-Euler system given in \cite{hF98}, with the hope that
particular features of this system allows enough decrease in the
differentiability threshold on the initial data to include the data
given here. A second way is to set up two initial boundary value
formulations, one for the interior of the body and one for the
exterior, and then match both, in an appropriate way, at the boundary
of the body (see \cite{hFgN99} and \cite{jmS98}). It is far from
being clear if either of these alternatives works.

\vspace{0.5cm}

{\bf Acknowledgments:} We wish to thank Helmut Friedrich for
suggesting this problem, Robert Beig and Marc Mars for nice
discussions, and Alan Rendall, Oscar Reula, Bernd Schmidt and Jeffrey
Winicour for reading the manuscript and suggesting several
improvements.  We also want to thank the first anonymous referee for
helpful comments and specially the second anonymous referee for
pointing out a mistake in the original manuscript and for many
illuminating suggestions.  G.N. was supported in part by a grant from
R\'egion Centre, France. G.N. also thanks the friendly hospitality of
the Relativity Group at The Enrico Fermi Institute, at The University
of Chicago, where part of this work was done.

\appendix
\section{Appendix}
\label{s-a}

We discuss here the bound on the physical energy density given by Eq.
(\ref{eq:b-vtheta}) in Theorem \ref{T:Hc-exist1}. In the first
subsection, we show an inequality that is true for all maximal,
asymptotically flat initial data with matter sources. This inequality
is similar to Eq. (\ref{eq:b-vtheta}) in the sense that relates the
same quantities, but only the $L^1(\Omega)$ norm of the energy
density appears. In the second subsection we show that
(\ref{eq:b-vtheta}) is in fact a restriction, that is, there exist
solutions of the constraint equations which do not satisfy it.
Nevertheless we give arguments to show that physical systems like
neutron stars do satisfy the bound (\ref{eq:b-vtheta}).

\subsection{An inequality}\label{s-tc}

Consider the following result.
\begin{lemma}
\label{L:gen-bound}
Let $M$, $\tilde M$, and $q_{ab}$ be as in Sec. \ref{S:MR}. Let
$p^{ab}$ and $\theta$ be any solution of
(\ref{eq:constr-m})-(\ref{eq:bc-h}) with $p_a{}^a =0$. Fix a point
$p\in \tilde M$, and denote by $B_r$ an open ball  centered at $p$,
of geodesic radius $r$. Then, for a sufficient small  $r$, we have
\begin{equation}
\label{eq:gen-bound}
\|\tilde \mu\|_{L^1(B_r)} 
\leq \frac{4^6}{5^5} \frac{2\pi r}{\kappa(\G_{-})^4} 
\end{equation}
where $\G_{-} := inf_{\partial B_r} \G$, where $\G$ is defined in
Eq. (\ref{eq:5}).
\end{lemma}
\goodbreak

\begin{proof}
Consider any solution, $\theta$ of Eq. (\ref{eq:constr-h}) on $\tilde
M$. Then,
\begin{equation}
\label{eq:muq}
\frac{\kappa}{4} \;\tilde \mu \leq - \frac{L_q(\theta)}{\theta^5}.
\end{equation}
Introduce $\vartheta$ as in Subsection \ref{s-exist}, that is, $\theta
= \G+\vartheta$. We parametrize all possible solutions, instead by
$\vartheta$, by a function $\sigma := L_q(\vartheta)$. Denoting by
$L_q^{-1}(\sigma)(x) =-(1/4\pi)\int_M \sigma(y) \G(x-y)\, dV(y)$, then,
inequality above translates into
\[
\frac{\kappa}{4} \;\tilde \mu \leq 
- \frac{\sigma}{[\G+L_q^{-1}(\sigma)]^5}.
\]
The Green function has the form, $\G(x-y) = 1/|x-y| +g$, where $g \geq
0$ on $B_r$.  (See \cite{Lee87,Schoen94}.)  The inequality
$\G(x-y) \geq 1/(2r)$, that is true for all $x,y \in B_r$, implies that
$L_q^{-1}(\sigma) \geq \|\sigma\|_{L^1(B_r)} /(8\pi r)$, and this,
in turn, implies
\[
\frac{\kappa}{4} \|\tilde \mu\|_{L^1(B_r)} 
\leq \frac{ \|\sigma\|_{L^1(B_r)}}{[\G_{-} 
+  \|\sigma\|_{L^1(B_r)}/(8\pi r)]^5}.
\]
The last step of the proof is to maximize the right hand side of
inequality above with respect to all possible functions $\sigma$. The
maximum value is taken for $\|\sigma\|_{L^1(B_r)} = 2\pi r\G_{-}$,
and inequality above gives Eq. (\ref{eq:gen-bound}).
\end{proof}

\subsection{Static spherical body}
\label{s-sp}

In this subsection we explicitly construct an initial data set for a
static, spherically symmetric, liquid-type body. We match, in
appropriate coordinates and in a $C^1$ way, a 3-sphere endowed with
its standard metric, with a 3-dimensional Schwarzschild slice. The
reason for re-doing this known construction (see \cite{Murchadha86b})
is twofold. First, this example, for suitable choices of the
parameters, violates the bound (\ref{eq:b-vtheta}).  Second, we want
to answer the following question: What kind of physical systems
satisfy the bound (\ref{eq:b-vtheta})? We show that the answer turns
out to be (at least for this example): stars with radius $R\geq 1.08
R_s$, where $R_s=2m$ is the Schwarzschild radius and $m$ is the total
mass. Note that this bound is bellow to $R\geq \frac{9}{8} R_s$, which 
is the necessary condition for hydrostatic equilibrium in General
Relativity (see for example \cite{Wald84}), then this bound is
expected to be satisfied for every star near equilibrium.

Let $M =S^3$, the conformal metric $q^0_{ab}$ be the standard metric,
unit radius, of $S^3$, the point $i$ be the North Pole of $S^3$, and
the domain $\Omega$ be a ball centered at the South Pole of $S^3$.

Let $\delta_{ab}$ be the flat metric, and $r$ be the corresponding
spherical radius.  Consider the following initial data set
 \begin{equation}
  \label{eq:tqd}
  \tilde q_{ab}=\hat \theta^4 \delta_{ab}, \quad \tilde p^{ab}=0.
\end{equation}
The conformal factor $\hat \theta$ is given by
\begin{equation}
  \label{eq:thetas}
\hat \theta=\begin{cases}
1+\frac{m}{2r} &\text{ if }  r\geq r_0,\\
a^{1/4}\left(\frac{2r_1}{r_1^2+r^2}\right)^{1/2} 
&\text{ if }  r \leq r_0,
\end{cases}
\end{equation}
where the positive constants $a$, $r_0$, $r_1$, and $m$, satisfy the
following relations 
\begin{equation}
  \label{eq:relations}
  r_1^2=\frac{2r_0^3}{m}, \quad a=\frac{(r_0+m/2)^6}{2mr_0^3}.
\end{equation}
Using \eqref{eq:relations} one check that $\hat\theta$, and hence
$\tilde q_{ab}$, is a $C^1$ function in $\mathbb{R}^3$.  There are
two free parameters, for example we can take $m$ and $r_0$, $m$ is the 
total mass of the data.  The 
metric $\tilde q_{ab}$  for $ r\geq r_0$ is the Schwarzschild  metric
in isotropic coordinates and for $ r\leq r_0$ is the standard metric
on $S^3$ of radius $a^{1/2}$. The Ricci scalar of the metric $\tilde
q_{ab}$ is given by
\begin{equation}
  \label{eq:Rsc}
  \tilde R=\begin{cases}
0 &\text{ if }  r > r_0,\\
\frac{6}{a}  &\text{ if }  r \leq r_0,
\end{cases}
\end{equation}
and the physical energy density is
\begin{equation}
  \label{eq:mu2}
  \tilde \mu=\begin{cases}
0 &\text{ if }  r > r_0,\\
\frac{6mr_0^3}{\kappa (r_0+m/2)^6}  &\text{ if }  r \leq r_0.
\end{cases}
\end{equation}
We see that the energy density $\tilde \mu$ has support in a closed
ball of radius $r_0$. In order to make contact with the assumptions
in theorem \ref{T:Hc-exist1}  we write this initial data as follows. 
Using the well known relation
\begin{equation}
 \label{eq:q0}
\G^4q_{ab}^0 = \delta_{ab}, \quad \G=
\left(\frac{r_1^2+r^2}{2r_1}\right)^{1/2},
\end{equation}
we obtain
\[
\tilde q_{ab} =  \theta ^4 q^0_{ab}, \quad  \theta =\hat \theta \G.
\]
For convenience, we have chosen a different normalization for the
Green function $\G$ than (\ref{eq:5}) and (\ref{eq:6}), in order to
fix $q^0_{ab}$ to be exactly the unit radius standard metric of $S^3$.
This difference in the normalization will, of course, play no role in
what follows.

We want to prove  that for some choices of the free parameters
$r_0$ and $m$, this initial data violate the bound
\eqref{eq:b-vtheta}. In order to do that we calculate explicitly the
right and left hand side of \eqref{eq:b-vtheta}. 
 Since  $\tilde \mu$  is constant
we have 
\begin{equation}
  \label{eq:2b}
||\tilde \mu ||_{L^2(\Omega)}
=\tilde \mu [\mbox{Vol}_{q^0}(\Omega)]^{1/2},
\end{equation}
where $\mbox{Vol}_{q^0}(\Omega)$ denotes the volume with respect to
the metric $q^0_{ab}$. 
From Eq. (\ref{eq:q0}) we have 
\begin{equation}
  \label{eq:3}
  \G_+=\left(\frac{r_1^2+r_0^2}{2r_1}\right)^{1/2}.
\end{equation}
Since $C_p=0$ in this example, using (\ref{eq:2b}) and (\ref{eq:3}) we 
obtain that inequality \eqref{eq:b-vtheta} is equivalent to 
\begin{equation}
  \label{eq:19}
  [\mbox{Vol}_{q^0}(\Omega)]^{1/2} \leq \beta (1+\frac{m}{2r_0})^4,
\end{equation}
where $\beta=4^5/(3\cdot5^5k) \approx 0.77$ (the constant $k$, which
depends only on $S^3$ and $q^0_{ab}$,  can be calculated
explicitly for this case $k=\sqrt{2}$). Note that
$\mbox{Vol}_{q^0}(\Omega)$ depends only on the dimensionless parameter 
$m_0/(2r_0)$. For $m_0/(2r_0)=0$ we have  that
$\mbox{Vol}_{q^0}(\Omega)=\mbox{Vol}_{q^0}(S^3)=2\pi^2> \beta^2$,
then there exist values of $m_0/(2r_0)$ such that the bound
\eqref{eq:b-vtheta} is not satisfied. We use that
$\mbox{Vol}_{q^0}(\Omega)\leq \mbox{Vol}_{q^0}(S^3)$ for arbitrary
$m_0/(2r_0)$, to obtain a sufficient condition in order to satisfy
Eq. (\ref{eq:19})
\begin{equation}
  \label{eq:20}
  m/(2r_0) \geq 1.75.
\end{equation}
Since the exterior metric is the Schwarzschild metric with mass $m$ we 
can write this condition in terms of the physical radial area
coordinate  $R$ to obtain
\begin{equation}
  \label{eq:21}
  R\geq 2.16 \, m.
\end{equation}


\end{document}